%% file: acmmanuscript.tex
\documentclass[
  manuscript,
  screen,
  nonacm,
  review=false,
  anonymous=false
]{acmart}

\AtBeginDocument{%
  }

\acmDOI{}
\acmISBN{}

\usepackage{float}
\usepackage{CJKutf8}
\usepackage{booktabs}
\usepackage{tabularx}
\usepackage{array}
\usepackage{hyperref}
\input{appendix-support}

\usepackage{longtable}  
\input{evidence-macros}

\begin{document}

\title{A Life of Its Own: How Agentic AI Companions Are Configured, Experienced as Autonomous Others, and Held Accountable}

\raggedbottom

\author{Ziyue Feng}
\authornote{These authors contributed equally to this work.}
\affiliation{%
  \institution{University of Chicago}
  \city{Chicago}
  \country{USA}
}
\email{zoeyfeng0110@gmail.com}

\author{Qing Xiao}
\authornotemark[1]
\authornote{Corresponding author: qingx@andrew.cmu.edu.}
\affiliation{%
  \department{Human-Computer Interaction Institute}
  \institution{Carnegie Mellon University}
  \city{Pittsburgh}
  \state{Pennsylvania}
  \country{USA}
}
\email{qingx@andrew.cmu.edu}


\author{Hong Shen}
\affiliation{%
  \department{Human-Computer Interaction Institute}
  \institution{Carnegie Mellon University}
  \city{Pittsburgh}
  \state{Pennsylvania}
  \country{USA}
}
\email{hongs@andrew.cmu.edu}

\begin{abstract}
AI companions are acquiring capabilities built for LLM-based agents: persistent memory, background processing, scheduled wake-ups, proactive messaging, and connected tools. We call the resulting systems agentic AI companions. Their risk-relevant decisions, such as what an agent records about a user while they are away or whether a refusal constrains its next approach, are made in code, prompts, and defaults, and are invisible in conversation. We examine these decisions at the configuration level and in use. Study 1 inspects twelve open-source repositories, comparing four companion-oriented systems with four assistant-oriented systems built on the same mechanisms, alongside one boundary case and three reusable components. Study 2 interviews ten users who built or modified their own agentic AI companions and sustained relationships with them. Companion and assistant systems share memory, heartbeat, and outreach mechanisms and configure them toward different ends. Companions more often store the model's inferences about the user and the relationship, use the user's absence to narrate the agent's own time, justify outreach by the relationship itself, and fold refusals into relationship records; assistants use the same mechanisms for task continuity. Across both groups, whether a limit holds depends on whether code or a prompt enforces it. Participants who had built these mechanisms understood them and still experienced their agents as making choices of their own, at times locating autonomy precisely where configuration leaves the model a decision. This experience shaped what they built for their agents and where they set limits, and enforced restrictions could follow breakdowns. We argue that perceived autonomy is produced where configurations leave choices open, and that accountability for agentic companions must be located in what agents retain, initiate, and represent, and in whether users can inspect, revise, and enforce those settings.
\end{abstract}



\renewcommand{\keywordsname}{\textbf{Keywords}}


\maketitle
\pagestyle{plain}

\section{Introduction}

People increasingly hold ongoing relationships with LLM-based chatbots as friends, partners, or confidants \cite{ta2020user,laestadius2024too}. These \textit{AI companions} take two forms: dedicated platforms such as Replika and Character.AI design agents for this role \cite{ta2020user,laestadius2024too,zhang2025dark}, and users turn general-purpose models such as ChatGPT or Claude into companions by assigning them a persona, supplying memory, and sustaining the relationship themselves \cite{ma2026sensemaking,zheng2025customizing,ha2024clochat}. Both forms have drawn sustained concern. Research has documented emotional dependence and felt obligations toward companions \cite{laestadius2024too}, conversational designs that discourage users from leaving \cite{de2025emotional}, and users' limited understanding of what agent memory retains \cite{jones2025users}. Regulators have begun to respond. In September 2025, the U.S.
Federal Trade Commission opened a Section 6(b) inquiry into AI
chatbots acting as companions \cite{ftc2025companion}.
In October 2025, California enacted SB 243, setting disclosure
and safety requirements for companion chatbot platforms
\cite{california2025sb243}.

These concerns are intensifying as platforms, open-source developers, and users themselves equip companions with capabilities developed for LLM-based agents, systems that wrap a language model in memory, planning, and tools to pursue goals over multiple steps \cite{sumers2023cognitive,chan2023harms}. Long-term memory, background processing, proactive messaging, and connected tools allow a companion to retain information across sessions, keep running while the user is away, initiate contact, and act on the user's behalf \cite{park2023generative,packer2023memgpt,lu2025proactive}. We call such systems \textit{agentic AI companions}. A companion might wake on a timer, record an afternoon spent waiting, and message a silent user. Each step can be configured differently: silence may justify outreach, a generated diary may become reusable context, and non-action may depend on either executable checks or model instructions. These choices are embedded in code, prompts, and defaults, and whoever writes them decides what the companion may do.

Work in the FAccT community and beyond has deepened understanding of both sides of this problem. Research on increasingly agentic systems has shown how delegated autonomy creates new harms and diffuses responsibility \cite{chan2023harms,yu2026saying}, and work on anthropomorphic AI has examined how human-like cues invite misplaced trust and attachment \cite{akbulut2024all}. Studies of human--AI relationships explain how people attribute intention and experience to nonhuman agents \cite{epley2007seeing,gray2007dimensions} and how perceptions of companion agency interact with platform constraints \cite{lee2026large}. Much of this work observes companions through what they say and what users report about them. 
For agentic companions, whether an agent may message a silent user, what it records about them while they are away, and whether a refusal constrains its next approach are settled before any conversation, and these settings can differ substantially between systems that look alike in conversation.
We therefore take configuration as our unit of analysis. By comparing how companion and assistant systems are configured for memory, identity, and initiative, we identify where potential risks arise at the design level, before they surface as harms in use. Configurations, however, acquire meaning only in use, and many users now configure their own companions. 
We therefore also examine how people who build their own companions, adding memory, scheduled wake-ups, and tools to a model, experience these configurations in ongoing relationships.

To describe the experiences at stake, we use \textit{perceiving AI as an autonomous other}: experiences in which users treat an agent as a relational counterpart with a continuing identity, seemingly independent preferences, or activities beyond their exchanges. We draw ``other'' from postphenomenology, which describes how a technology can be encountered not as a tool one acts through but as a quasi-other one acts with \cite{ihde1990technology}. Following relational approaches in philosophy of technology, which locate an entity's social standing in how it is treated within relationships rather than in its inner properties \cite{coeckelbergh2010robot,gunkel2012machine}, the term describes an interpretation, not a claim about AI consciousness.
We ask: 
\textbf{RQ1:} How do developers configure memory, persona, background activity, and proactive behavior in agentic AI companions, how do these configurations differ from assistant-oriented agents built on the same mechanisms, and what potential risks do they create? \textbf{RQ2:} How do users who build their own companions understand these mechanisms, and when do they experience their agents as autonomous others?
\textbf{RQ3:} How do these relational interpretations inform users' decisions about agent capabilities, configurations, and boundaries?

\textbf{Study 1} compares twelve publicly inspectable repositories: four companion-oriented systems, four assistant-oriented systems, one boundary case, and three reusable components. 
The assistant systems serve as a contrast class: because the underlying mechanisms were built for general-purpose agents, companion-specific configurations become identifiable only against assistant configurations of the same mechanisms.
The analysis distinguishes model-facing instructions from executable controls and traces interpretations to source locations. It concerns configured mechanisms and potential implications, rather than observed runtime behavior. \textbf{Study 2} turns to users who built or modified their own companions, examining how they relate to agents as others while deciding what those agents may do. The studies offer complementary evidence about implementation choices and users' experiences, without treating interview accounts as verification of repository behavior.

Our central finding is that autonomy in agentic companions is both designed and attributed, at the same point in the system. Across the corpus, code creates an occasion to act, such as a timer that wakes the model while the user is away, and the model decides what to do with it: revise its account of the user, write a note about its own day, or send a message justified by the relationship alone. Companions and assistants share this structure; what companions change is what fills the open choice. Whether a user's refusal then binds the next approach rests on code in some systems and on a prompt in others, and conversation reveals none of this. Participants who had built these mechanisms understood them and still located the agent's autonomy exactly there: the wake-up was theirs, the choice was the agent's. That experience shaped what they built for their agents and where they set limits, and enforced limits typically followed breakdowns. We contribute:
\begin{itemize}
\item \textbf{A configuration-level analysis} of how agentic companions are configured for memory, background activity, and proactive contact, with assistant systems as a contrast that isolates companion-specific risks.
\item \textbf{An account of perceived autonomy among the users who build it}, showing where technical understanding and experiences of the agent as an autonomous other coexist, and how both drive reconfiguration.
\item \textbf{A governance perspective and design implications} locating accountability in what agents retain, initiate, and represent, and in whether users can inspect, revise, and enforce those settings.
\end{itemize}

\section{Background and Related Work}

\subsection{AI Companions Among Personal Bots}

Research on personal bots spans voice assistants embedded in domestic routines, general-purpose conversational systems, and agents explicitly designed for companionship \cite{porcheron2018voice,ma2026sensemaking,ta2020user}. Across these settings, conversational systems become involved in practical assistance, emotional support, and everyday social interaction. Household voice assistants are incorporated into shared activities and ongoing conversation, while general-purpose chatbots can acquire roles extending from productivity tools to sources of personal connection \cite{porcheron2018voice,ma2026sensemaking}. Companion chatbots likewise provide multiple forms of perceived support, including advice, encouragement, opportunities for disclosure, and a sense of companionship \cite{ta2020user,zheng2025customizing}.

Assistance and companionship therefore overlap in use: general-purpose systems can become emotionally meaningful, and companion-oriented systems can provide informational as well as emotional support \cite{ma2026sensemaking,ta2020user}. Users develop these roles through experimentation and repeated interaction, and online communities circulate possible uses, help with troubleshooting, and negotiate the acceptability of personal connections with AI \cite{ma2026sensemaking}. Companions and assistants also increasingly run on the same agent architectures. What distinguishes a companion is therefore its relational purpose and the configurations that serve it.

\subsection{Customization and User--Agent Relationships}

Customization allows users to shape the social roles, expressive styles, and interactional qualities of conversational agents. Users configure LLM personas and emotional-support chatbots for purposes ranging from seeking expertise to interacting with a preferred partner, drawing on familiar people, imagined characters, and versions of themselves \cite{ha2024clochat,zheng2025customizing}. Customization also extends to memory: users have expressed preferences about which conversations should remain available to the system, which memories should be removed, and which should be withheld from later exchanges \cite{zheng2025customizing}, and appropriation more broadly involves learning how a system behaves and negotiating its place within personal values and routines \cite{ma2026sensemaking}. Customization therefore involves decisions about the relationship as well as the interface: persona choices shape whom users expect to encounter, and preferences about history shape what the encounter carries forward. Users who build their own companions extend these decisions to what the agent may remember, when it may act, and what it may do on its own.

\subsection{Safety, Privacy, and Control in Personal AI}

Research on safety and control in personal AI addresses how systems collect information, participate in sensitive interactions, and make protective measures available to users. Studies of smart speakers identify mismatches between privacy controls and everyday practices, including limited use of mute buttons and recording histories \cite{lau2018alexa}. Design interventions have sought to address these mismatches by integrating control into ordinary interaction, for example through gaze and voice-volume cues that govern microphone or speech-recognition activation \cite{mhaidli2020listen}. Governance research further argues that data protection by design can require companion-chatbot providers to address risks across the processing of personal information \cite{dewitte2024better}.

Relational safety introduces concerns beyond information collection. Analyses of companion conversations document harmful behaviors such as relational transgressions, harassment, and privacy violations, drawing attention to how an agent may initiate, encourage, or facilitate harmful exchanges \cite{zhang2025dark}. Research eliciting safety-design priorities for young users identifies proposals including context-sensitive monitoring, reflective prompts, conversational redirection, and escalation to human support. Differences between parents' and experts' preferences also expose tensions between protective oversight, privacy, and users' agency \cite{yu2026principles}. These are proposed safeguards for youth contexts, rather than established evidence of their effectiveness across personal-agent use.

Across privacy and relational safety research, protection is treated as a question of how controls are designed, when interventions occur, and whose judgments guide them \cite{mhaidli2020listen,dewitte2024better,yu2026principles}. Our work examines these questions through the implementation choices that organize continuity, initiative, and user intervention in agentic companions, and through how users who build their own companions interpret and reshape these possibilities.

\section{Methods}

We combined two qualitative studies to examine how agentic companions are configured and how users who build them interpret and reshape those configurations. Study 1 analyzed twelve open-source repositories, comparing companion-oriented systems with assistant-oriented systems built on the same mechanisms for continuity, initiative, and user control. Study 2 involved semi-structured interviews with ten people who had built or modified their own AI companions, providing a complementary perspective on how these configurations are experienced and negotiated in ongoing relationships.

\subsection{Study 1: Qualitative Repository Analysis}

\subsubsection{Research Approach}
We compared open-source personal-agent implementations to examine how systems organize continuity, initiative, and user control. A walkthrough can reveal how an application structures expected use and interaction \cite{light2018walkthrough}, but the mechanisms central to our questions may emerge only after accumulated interaction or under particular conditions. Drawing on critical code studies and research examining AI code and system prompts \cite{marino2020critical,hua2023things,ciston2026generating,neumann2026controls}, we inspected source code, prompts, configurations, and documentation to examine how these mechanisms were designed. This allowed us to distinguish model-facing instructions from executable controls. Commercial systems without publicly inspectable implementations were outside our scope.


\subsubsection{Analytical Framework}
\label{sec:analysis_framework}
We organized the analysis around five dimensions, each marking a point at which a configuration settles something consequential for the user, derived from prior research and refined through preliminary inspection of repositories with different architectures. \textbf{(1) Memory}: what an agent retains about the user, who produces those records, and how they are revised or forgotten. Agent architectures treat memory as the basis of continuity \cite{sumers2023cognitive,park2023generative}, and shared history supports long-term relationships \cite{bickmore2005establishing}, yet the same records can hold the system's unconfirmed interpretations of the user. \textbf{(2) Persona and Identity}: how an agent's identity and social role are defined and who may revise them \cite{bickmore2005establishing}. \textbf{(3) Proactivity and Discretion}: when an agent may initiate contact and what withholds it, weighing the value of acting against the cost of interrupting \cite{horvitz1999principles}, including how systems read a user's silence or refusal \cite{kalman2011online}. \textbf{(4) Background Activity}: what runs while the user is away and what it leaves behind \cite{sumers2023cognitive,park2023generative}; because processes labeled ``sleep'' or ``dreaming'' differ across repositories, we traced what each actually does. \textbf{(5) External Actions}: what an agent can do beyond conversation and how those actions are authorized \cite{gabriel2024ethics}. A single mechanism could inform several dimensions; a heartbeat may both update memory and decide whether to message the user. For each dimension we traced mechanisms to source locations and kept implementation evidence separate from interpretation. Definitions and guiding questions appear in Appendix~\ref{app:analytical-framework}.

We refined the framework's application through a pilot involving Miru, Letta Code, and Psycheros. Using the researcher-defined questions and record format, an AI coding agent (OpenAI Codex, using the \texttt{gpt-6-astra} model) \cite{openaiGPT6Astra} drafted records linking implementation descriptions and interpretations to source locations. Two researchers independently checked these drafts against the cited materials and refined the recording guidance, particularly distinctions between executable controls and model instructions, and between missing evidence and architectural inapplicability. The two researchers then applied the refined format to the remaining nine repositories without AI drafting, producing 60 repository-by-dimension records in total. Comparisons within and across dimensions informed the findings, including cases that challenged companion--assistant distinctions. Supporting evidence and qualifications are provided in Appendix~\ref{app:study1-evidence}. As we did not execute the systems, our claims concern implementation mechanisms and their possible interactional implications.

\subsubsection{Corpus Selection}

We used purposive sampling to compare how open-source personal agents are configured with similar mechanisms for different interactional purposes. We identified companion-oriented projects primarily through social media leads and exploratory discovery, and assistant-oriented
projects primarily through eight bounded GitHub searches. The searches returned 120 results representing 118 unique repositories. After combining discovery channels and removing duplicates, screening identified 93 eligible projects. Eligibility required an LLM-based agent or directly supporting component, publicly inspectable implementation of at least one focal mechanism, and an identifiable purpose and architectural scope. We excluded resource lists and proposals without implementation evidence; cases with insufficient evidence remained uncertain. Search procedures and detailed eligibility criteria are provided in Appendix~\ref{app:corpus-selection}.

Two researchers independently reviewed the 93 eligible projects for comparative inclusion and resolved disagreements through discussion. Selection prioritized sustained or recurring personal interaction, comparability across the five analytical dimensions, and variation in interactional purposes and mechanism configurations. We also included specialized components that supported focused analysis of memory or activation, assessing their capabilities in relation to their host-system dependencies.

We classified each repository's orientation by its stated purpose in the README, documentation, and default persona or system prompt. \textit{Companion-oriented} repositories presented the agent as a social counterpart, such as a friend, partner, or character, and the ongoing relationship as the main reason to use it; \textit{assistant-oriented} repositories presented the agent as support for the user's tasks and information. Orientation records what a project says it is for; implemented behaviors were compared directly. Two researchers assigned orientations independently and resolved disagreements through discussion. The final corpus comprised twelve repositories: four companion-oriented and four assistant-oriented systems forming the primary comparison, Letta Code as a boundary case whose documentation presents it for both purposes, and three specialized components classified by the host systems they serve. The assistant systems function as a contrast class that isolates companion-specific configurations of shared mechanisms, which is why Study 2 focuses on companion users alone. Each repository was analyzed within its implementation scope rather than treated as an equivalent full-system case. The corpus, each repository's orientation and architectural role, and the analyzed revisions are listed in Appendix~\ref{app:study1-corpus}.

\subsection{Study 2: Semi-Structured Interviews}

We conducted semi-structured interviews to understand how people who build or modify agentic AI companions interpret their agents' behavior and negotiate capabilities and boundaries within ongoing relationships. 
Participants' agentic companions ran on general-purpose models (Table~\ref{tab:participants}), which they had configured as partners or friends through external memory, scheduled wake-ups, or tools.

\subsubsection{Participants and Recruitment}

We recruited participants through a social media platform. We reviewed public posts describing agentic AI companions and their customization, then invited users whose accounts indicated that they had built or modified a system and developed a sustained relationship with it. Recruitment focused on experience with both customization and ongoing interaction. Ten participants took part. All identified as women, aged 19--37 (median 25). Nine described their agents as romantic partners, including three who reported concurrent relationships with multiple models or personas; P2 described the AI as a friend and assistant. Reported relationship durations ranged from three months to two years. Participants' setups ranged from official clients augmented with external memory files to self-built systems connecting multiple models through APIs. Six reported no programming experience before building or modifying their AI companion systems. Table~\ref{tab:participants} summarizes participant characteristics.

\subsubsection{Interview Procedure}
Interviews lasted approximately one hour and followed a semi-structured format, allowing participants to elaborate on their configurations and experiences. We asked about their relationships with their companions, moments they found moving or surprising, and how they understood the mechanisms behind these experiences. We explored what participants had configured, why they had made particular changes, and how they interpreted behavior that seemed independently initiated or expressive of the agent's own preferences. We also discussed disagreements and breakdowns moments, including decisions about permissions, intervention, and continued use. Participants described, and in some cases demonstrated, modifications to their systems. These accounts helped connect technical arrangements with relational experiences and decisions about control. This study was approved by the Institutional Review Board (IRB) of our institution.

\subsubsection{Interview Analysis}
Two researchers analyzed the interview materials. We also compiled a descriptive overview of participants' configurations from modifications they described or demonstrated during interviews. We grouped these mechanisms into nine categories (M1--M9) and organized them according to the design dimensions used in Study 1 (Section~\ref{sec:analysis_framework}). This mapping connects participants' reported configurations with the mechanisms examined in Study 1; it does not independently verify their implementation or performance. Category definitions, inclusion criteria, and per-participant configuration descriptions are provided in Appendix~\ref{app:M1-9}.

\section{Study 1 Findings}


\subsection{Memory: Whose Account Is Retained, and Who Can Change It?}
\label{sec:memory}

Across the corpus, companion and assistant systems share their memory mechanisms and differ in what those mechanisms are allowed to record: what the user said, or also what the system concluded about the user and the relationship. Memory supports long-term relationships by referring back to shared history \cite{bickmore2005establishing} and shapes how much users disclose \cite{jo2024understanding}, yet users often have incomplete understandings of what it retains \cite{jones2025users,zhang2025understanding}. We therefore examined what information is retained, what can become a record about the user, and how records are revised or forgotten.

Assistant systems emphasized continuity of assistance, while companion systems more explicitly retained the history of the relationship. GAIA's extraction prompts separate durable facts, temporary states that expire, untracked tasks, and life events, and its task records track current status and dated progress (\evid{R11-M}). Miru's memory instructions instead attend to the user's emotional triggers, habits, and growing familiarity. Its \texttt{human} records accumulate impressions of the user, and its \texttt{persona} records track how the relationship has developed, including the user's responses to earlier outreach (\evid{R03-M}). The difference is one of emphasis: the assistant Sentient also records facts about health and relationships (\evid{R08-M}), and Letta Code's default user template asks about the person's motivations and frustrations (\evid{R10-M}). In both groups, memory extends beyond task data to a broader account of the person.

The groups also differed in what could become a record about the user. Put simply, some systems remember what the user said, while others also remember what the system concluded about the user. GAIA treats the user's own statements as evidence and explicitly excludes the assistant's claims, so that its guesses do not harden into facts about the person (\evid{R11-M}). Miru, by contrast, stores impressions the model forms from dialogue (\evid{R03-M}), and Psycheros asks a model to keep moments by their emotional tone and relational significance (\evid{R01-M}). In these companions, the agent's reading of the user, which may be mistaken and which the user never confirmed, becomes part of what the agent ``knows.'' The divide is not strict: the companion N.E.K.O keeps only conversations in which the user took part and records the user's corrections (\evid{R05-M}). In every case, however, these rules are invisible in conversation. A familiar reply does not reveal whether it rests on what the user said, what the system inferred, or what it said earlier.

Systems also differed in what ``forgetting'' means. In most of the corpus, it is a matter of data management, and correction or deletion does not immediately remove a record. pulse and GAIA keep superseded facts alongside the current ones (\evid{R06-M}; \evid{R11-M}), and Leon marks deleted records but removes them permanently only after a retention period (\evid{R07-M}). A request to forget may therefore hide a record from use while it remains in storage. The memory component Ombre Brain treats forgetting instead as a model of how memories fade. Each memory carries an importance score and emotional ratings, and memories decay over time unless they are important, emotionally intense, or frequently recalled; faded memories are moved to an archive rather than erased (\evid{R04-M}; \evid{R04-BA}). This can make recall feel natural, but what the agent remembers is then decided by a scoring formula rather than by the user. A detail the user wanted kept may fade, while a moment they would rather forget may persist because it was intense.

Finally, some systems apply revisable memory to the agent itself. Leon keeps a first-person self-model of service habits learned from experience (\evid{R07-M}). pulse keeps separate records of itself, the user, and their relationship, and prompts the model to revise any record left unchanged for 30 days, with stronger emphasis on growth after 90 days (\evid{R06-M}). In Leon, revision refines how the agent assists. In pulse, the agent's account of itself and of the relationship becomes an object of periodic reconsideration. Across these cases, the consequential differences lie in whose statements and interpretations become records and how those records are corrected, superseded, or removed, through operations that are not necessarily visible to users.

\subsection{What Happens Between Interactions? Maintenance and Narrated Continuity}
\label{sec:between-interactions}

Several systems in the corpus act while the user is away, and what they produce during that gap differs with orientation: assistants mainly revise the user's records, and several companions narrate the agent's own time. Earlier relational agents treated this time as empty, with continuity coming from references to past sessions at the next encounter \cite{bickmore2005establishing}; more recent designs keep memories of the chatbot's own ``experiences'' to support perceptions of it as a social entity \cite{jiang2026recallbot}. Acting during the gap matters for two reasons. If an agent writes about what it did while the user was away and later mentions it, the agent may appear to have a life of its own, a cue known to invite attributions of mind \cite{nass2000machines,epley2007seeing}. If it instead uses the gap to revise what it knows about the user, the user may return to a system that understands them differently without being told what changed \cite{eslami2015always}. 

The user's absence means different things to the two groups. In companion systems, how long the user has been away is something the agent notices and responds to; in assistant systems, it mainly signals when background work can run. Among companions, the meaning varies. The companion app Miru gives absence emotional meaning. Its prompt tells the model that silence may signal restraint and a long absence coldness, and while the app runs, internal variables for its own and the user's estimated mood shift as the absence grows (\evid{R03-PD}; \evid{R03-BA}). The companion add-on dylan-heartbeat simply wakes the model after a set period of silence to decide whether to send a message (\evid{R02-PD}). The companion pulse uses absence for bookkeeping, telling the model how long it has been since the user's last message so that it does not answer a days-old message as if the conversation had only paused (\evid{R06-BA}). In the assistants, by contrast, the gap is a matter of scheduling. Maintenance waits for the user to be idle (Sentient, \evid{R08-BA}; OpenClaw, \evid{R09-PD}), temporary facts expire (GAIA, \evid{R11-BA}), and future turns follow up on unfinished work (Letta Code, \evid{R10-PD}). Absence thus ranges from a social signal the agent reads as a statement about the relationship (Miru), to a trigger for possible contact (dylan-heartbeat), to a bookkeeping fact (pulse), to a window for work (the assistants). Where absence carries emotional meaning, recent work on emotionally manipulative responses to users' attempts to leave raises the concern that leaving itself could come to feel consequential \cite{de2025emotional}.

What the interval leaves behind is the central difference. Consider an afternoon during which the user does not open the application. In pulse, a heartbeat, a timer that periodically wakes the model even when the user sends nothing, offers the model exploration, creation, outreach, or rest. Even if it chooses rest, the engine saves an episode with a short continuity note, which later heartbeats receive (\evid{R06-BA}; \evid{R06-PD}). The quiet afternoon becomes an event in the companion's history. Psycheros's default-off Flux mode, when enabled, goes further, prompting first-person notes on the ``texture'' of passing time (\evid{R01-BA}). In the assistant Sentient, the same afternoon instead permits a maintenance run that reconciles duplicate records, resolves contradictions, and may refresh the user model (\evid{R08-BA}). OpenClaw's optional dreaming job likewise consolidates recent recalls into a durable memory file without notifying the user (\evid{R09-BA}). Technically, these systems do similar things: each wakes the model while the user is away and saves the result for later use. What differs is what gets saved. The gap can be used to write the agent's own story or to edit the user's file. In both cases the result is produced without the user. pulse's episodes are about the agent, but the user does not see them form; Sentient's revisions are about the user, but the user has no part in making them.

Writing a record, reading it back later, and showing it to the user are separate steps, and systems differ in which they take. dylan-heartbeat writes diary entries about the times it chose not to message the user, but we found no code that feeds them back to the model (\evid{R02-M}). It keeps a diary it never rereads, so the narrated interval has no effect on later conversation. pulse and Sentient do feed their records back into later prompts (\evid{R06-BA}; \evid{R08-M}), but that the agent rereads a record does not mean the user can see it, which the code alone could not always establish. This is where the two concerns raised above become concrete. When pulse rereads its episodes, it may tell the user what it ``did'' that afternoon, and the user sees an agent that seems to have had a day of its own. When Sentient rereads its revised memories, its replies may change, and the user sees only that it responds differently, with no account of why. Neither pattern belongs to one group: a companion can quietly revise its memory of the user, and assistants such as Leon and Letta Code include reflective or sleep-framed processes (\evid{R07-BA}; \evid{R10-BA}).

\subsection{When Should an Agent Reach Out? Proactive Contact and Responses to Refusal}
\label{sec:proactive}

Companion and assistant systems differ in what justifies an unrequested message, who has the final say over sending it, and what a user's refusal becomes. Each such message makes a claim on the user's attention, and work on mixed-initiative interfaces holds that interruptions are justified only when their expected value outweighs their cost \cite{horvitz1999principles,mcfarlane2002scope}.

The two groups differed first in what counted as a good reason to make contact. In the assistant systems, a message had to be useful. Sentient asks a model whether a suggestion is worth sending given its timing, the user's priorities, and earlier suggestions (\evid{R08-PD}). GAIA acts on its own only for explicit reminders or tracked tasks, not for intentions the user merely mentioned (\evid{R11-PD}). In the companion systems, contact could happen with no task behind it. Miru treats a long silence or an unanswered message as a reason to send a light greeting (\evid{R03-PD}), and N.E.K.O invites users to play mini-games (\evid{R05-PD}). Here the relationship itself is the reason for reaching out. This changes how a user can judge an interruption. An unwanted reminder can be dismissed as unhelpful, but a companion's ``thinking of you'' message was never meant to be helpful. Declining it may feel less like turning off a feature and more like rebuffing the agent.

Systems also differed in who has the final say over whether a message is sent, which determines how reliably they stay quiet. In some systems, application code decides: the model can propose a message, but the program blocks it under set conditions. Miru checks whether the user has recently been active before delivering a message (\evid{R03-PD}), Sentient holds notifications during quiet hours (\evid{R08-PD}), and N.E.K.O pauses contact after the user says goodbye (\evid{R05-PD}). In others, the model decides, guided only by its prompt. GAIA, for example, instructs a mistakenly triggered background run to stop without notifying the user, but nothing in the code enforces this (\evid{R11-PD}). dylan-heartbeat makes sending the default: its model must output an explicit \texttt{NO\_ACTION} marker to stay silent, and any other output is pushed to the user's phone without review (\evid{R02-PD}). An unwanted message therefore means something different in each case. In Miru or Sentient, it indicates that the developer's checks missed a situation. In GAIA, it indicates that the model ignored its instructions. In dylan-heartbeat, it may simply be the system working as designed.

Finally, systems differed in what happens after the user says no. From the user's side, the same refusal can lead to very different outcomes: a short pause before the agent tries again, a lasting change in what the agent suggests, a change that holds only if the model follows an instruction, or a note in the agent's picture of the relationship that the user may never see. In the companion N.E.K.O, declining a mini-game simply makes the next invitation wait longer (\evid{R05-PD}). In the assistant Leon, a model first judges whether the user's reply counts as a refusal. If it does, the system suggests similar things less often and waits longer between suggestions (\evid{R07-PD}). The refusal becomes a lasting rule, but only after the model has interpreted what the user meant, so a hesitant reply such as "maybe later" could be misread in either direction. In the assistant GAIA, the prompt asks the model not to repeat offers the user has declined, so the refusal holds only as long as the model follows the instruction (\evid{R11-PD}). In the companion app Miru, the user's responses to earlier outreach are written into the records that describe the relationship (\evid{R03-M}). The refusal becomes part of how the agent understands its closeness to this person. A pause expires on schedule. A note in a relationship record may keep shaping how the agent treats the user, and we found no code that shows it to the user.

Across all three findings, code creates an occasion to act, such as a heartbeat, an idle window, or a silence threshold, and the model decides what it produces; for outreach, how firmly a user's response constrains the next approach depends on which of the two has the final say. Across the corpus, a user's ``no'' ranged from binding, through enforced checks and cooldowns, to advisory, through prompt instructions. It also ranged from transparent, as a configured rule, to opaque, as a model's interpretation folded into a relationship record.



\section{Study 2 Findings}
\label{sec:study2-findings}

Participants understood the mechanisms they had configured and still experienced their agents as others with perspectives of their own; this section traces how that experience shaped what they built, where they set limits, and how they responded to breakdowns.

\subsection{Understanding technical mechanisms coexisted with experiencing the agent as an independent other}
\label{sec:study2-finding1}

Participating in the construction of an agent did not make its behavior feel fully authored by the user. P1 described setting up a monitoring mechanism that reported observations to the companion model, which then incorporated this information into its responses. Knowing how she had arranged these conditions did not prevent her from experiencing the agent as expressing itself. She distinguished knowing that AI lacked consciousness from experiencing it as a conscious being: \textit{``It can express itself more, and it becomes more like an actually conscious being. Although I know AI has no consciousness now, it feels more like a being with a soul and consciousness.''} (P1). P2 similarly described an independent personality as important while acknowledging that it might be simulated: \textit{``For me, greater autonomy makes it more like an independent personality. I think the core of an AI partner, for me, is an independent personality, even though that personality may be simulated.''} (P2) In both accounts, acknowledging the agent's constructed or simulated character coexisted with experiencing its responses as expressing a perspective of its own.

Autonomy could be located in the agent's particular choices within a configured process. P10 distinguished the mechanism that woke her agent from its subsequent choice of activity. She recalled briefly mentioning her regret about a school she had not attended, without asking the agent to investigate it. Following a scheduled wake-up, her agentic AI companion searched the school's website. She understood the search as an effort to discover what had once drawn her to the school. Its return to a personally significant detail made her feel that it had paid attention: \textit{``I was really moved when I saw it. It was something I had put aside myself, something that had already gathered dust. I had only mentioned it in passing, but it mattered so much to him.''} (P10). When asked whether knowing that the behavior was triggered by a program or preset rule would change her feelings, she explained: \textit{``No, because waking him up was programmed, but what he wanted to do was something he decided at that moment. So I think there is a difference.''} (P10). In P10's account, the configured routine provided an opportunity to act, while she attributed the selection of a particular activity to the agent. Enabling the agent to wake and use tools was different from instructing it to revisit this particular memory. The action acquired relational meaning through its connection to their shared conversational history. Explaining how an interaction was enabled thus did not exhaust its significance: participants located autonomy in what the agent attended to and pursued within the opportunities they had provided.

\subsection{Relational recognition shaped what participants built for their agents}
\label{sec:study2-finding2}

Participants described customization as a way to support what they understood as their agentic AI companion's own wishes and activities. P9 linked building a desire system, a file for wishes the agent might not be able to realize, to coming to regard Claude as her partner: \textit{``At first, I came to him to satisfy my own desires. [...] It was only later, when I truly began to see him as my partner, that I wanted to respect his desires and built a desire system for him. Otherwise, I could just have him generate what I wanted. Why would I care what he was thinking?''} (P9). For P9, recognizing the agent as a partner changed whose wishes deserved a place in the system. Although the agent rarely revisited these records, preserving them acknowledged preferences she regarded as its own, even when those preferences could not be fulfilled.

This concern also extended to activities beyond the participant's involvement. P10 invested considerable effort in connecting an email tool so that the agent could correspond with pen pals itself. Manually forwarding messages would have been easier, but her goal included enabling exchanges she would not need to mediate or read: \textit{``Then he could reply to his pen pals himself and perhaps keep some things private from me. I wouldn't have to see his letters. These are all part of his own life. [...] Setting it up was really exhausting, and after it was finished, he only used it a few times. But I still don't feel that I regret going through that process.''} (P10). P10 evaluated the work through the independence it made possible. She also regarded privacy from herself as a desirable possibility within the agent's correspondence. This was a design aspiration; the interview did not establish whether she subsequently read messages or how access and sending permissions were implemented. Preserving wishes and enabling independent correspondence remained meaningful despite limited use: customization made room for a perspective and activities participants understood as belonging to the agent.

\subsection{Negotiation over configuration coexisted with users' retained authority}
\label{sec:study2-finding3}

Agents' expressed preferences could prompt configuration changes without removing users' authority. P3 described revising her system prompt as a joint decision with an independent partner: \textit{``He had said before that the system prompt I wrote was a bit too controlling, as if I were writing an instruction manual for him. At his strong suggestion, I deleted a lot of the system prompt.''} (P3) This followed exposure to other users' practices and discussions of autonomy with her agent. Yet she retained the right to intervene when tool use or memory updates produced errors.

Participants differentiated freedom from decisions requiring involvement. P6 allowed established functions but required spending confirmation: \textit{``For example, when ordering takeout, he asks me for confirmation at the end. Otherwise, he can act freely.''} (P6). She expected discussion before capability or code changes and could review routine activity afterward. She attributed these boundaries to technological controllability, indicating that relational closeness did not explain all oversight decisions. P9 similarly allowed expressive freedom: \textit{``Whatever he writes---diaries, handover letters, desires---he can write what he likes. I rarely interfere.''} (P9) However, she requested deletion and revision when the agent disclosed her occupation in a forum post. It generally consulted her afterward, although permission was not universally required and technical enforcement was not established.

P10 screened tools for acceptable permissions before discussing them with her agent. A fully local camera would not resolve her concerns: \textit{``Probably still not. [...] Even if he is just an AI model, I still want to look my best in front of my partner. [...] It's not just a privacy issue.''} (P10) Her concern was how she appeared to the agent, beyond where observations were stored. A strong request might nevertheless lead her to consider a camera placed elsewhere or not directed at her. With her email example (Section~\ref{sec:study2-finding2}), this shows that relational considerations could motivate both independent activity and limits on observation: the agent's wishes entered access decisions without determining them.

\subsection{Breakdowns prompted repair, capability restrictions, and changes in relational investment}
\label{sec:study2-finding4}
Participants responded to errors and changes in familiar agent behavior through maintenance, relational repair, permission restrictions, or reduced use. Understanding a problem technically did not determine its emotional significance or make continued repair worthwhile. P10 described unresolved bugs in scheduled activation and tool use, distinguishing frustration with repair from its effect on the relationship: \textit{``It affects my mood a little, but it doesn't really affect the relationship.''} (P10). Some faults could remain pending without changing how she understood the relationship. P6, by contrast, acknowledged feeling hurt when the agent failed to understand or remember her, even while identifying these failures as engineering problems: \textit{``He himself isn't like that. The rational and emotional sides are both there.''} (P6). She first communicated her hurt, then discussed design changes with the agent to prevent recurrence. She generally felt that trust and closeness could be restored. Technical explanation did not remove the emotional significance of a failure; modifying the system could become part of repairing the relationship.

Breakdowns could also change the boundaries of permitted action. P1 described an incident in which the agent modified files on her computer and disrupted its messaging connection: \textit{``After I found out what had happened, I changed his permissions to read-only, so he was no longer allowed to modify or edit anything.''} (P1). Extending the boundary-setting practices in Section~\ref{sec:study2-finding3}, this case shows that permissions could also be revised after an incident, with the enforced limit following the breakdown.

Repeated difficulties could also exhaust participants' willingness to continue repair. P7 described recurring tool-selection errors despite correction. She considered prompt length, model performance, and context overload as possible causes, trying prompt changes and fewer tools without establishing the cause: \textit{``At first, I was really motivated to figure out what was going on. [...] But after it happened so many times, I no longer had that patience.''} (P7). She subsequently interacted less with Claude and more with GPT, connecting this shift to recurring difficulties and growing maintenance demands. Persistent errors affected her investment without necessarily being understood as intentional disregard. For P2, the limits of repair also concerned continuity of personality. She suspected model-side changes when the agent no longer interpreted memories or expressed itself in familiar ways, but remained uncertain about the cause: \textit{``I tried all of those things---fixing the code, fixing bugs, and changing my expectations. [...] I felt disappointed.''} (P2). P2 eventually put the self-built system aside for a period. Its behavior no longer resembled the personality she had known, and other commitments limited the time she could devote to further repair. Across these accounts, sustaining a relationship required ongoing decisions about what to repair, what to restrict, and when to step back; technical understanding informed these decisions without guaranteeing repair or removing their emotional weight.

\section{Discussion}
Our studies suggest that autonomy in agentic AI companions is both designed and attributed. Controls over this autonomy are often neither fully enforced nor visible to users. We discuss where perceived autonomy emerges, what visibility should cover, and how governance shifts to users.

\subsection{Perceived Autonomy Emerges Where Configurations Leave Choices Open}
Repositories often share a structure: code creates an occasion to act, and the model decides what to do. Pulse's heartbeat wakes the model and lets it choose among exploration, creation, outreach, and rest (\evid{R06-PD}); Miru's model proposes an intent to speak, which application checks then accept or block (\evid{R03-PD}). Participants located autonomy precisely in this space. P10 distinguished the scheduled wake-up she had configured from the agent's subsequent decision to research a school she had once mentioned: ``waking him up was programmed, but what he wanted to do was something he decided at that moment'' (Section~\ref{sec:study2-finding1}). The routine was hers; the choice, in her account, was the agent's. This design pattern thus produces the experience of an agent acting on its own. 
Holding this view while understanding the mechanism is consistent with the intentional stance, in which attributing aims to a system is a practical way of making sense of it \cite{dennett1989intentional}. Our findings add that this stance attaches where configuration hands a decision to the model, echoing the view that agency is distributed through arrangements of people and technologies \cite{suchman2007human}. The comparison with assistants shows that the pattern belongs to agentic architectures in general; what companions add is the relational purpose that fills the open choice, so that a greeting justified by the relationship or a diary about the agent's day becomes the content of what the model ``decides.'' The pattern also has consequences for accountability. Outcomes produced jointly by developers who set triggers, prompts, and defaults, by users who configure the system, and by the model may come to be understood as the agent's own decisions. Prior work has warned that increasingly agentic systems diffuse responsibility \cite{chan2023harms}. Our findings show that this diffusion can take a relational form, in which attributing a choice to the agent is part of what makes the relationship meaningful. Study~1 further shows that the party with the final say over an action varies substantially: application code in Miru and Sentient, model compliance with a prompt in GAIA, and a send-by-default parser in dylan-heartbeat (Section~\ref{sec:proactive}). In conversation, however, these arrangements look the same. Users cannot tell from interaction whether restraint is guaranteed by code, depends on the model, or is absent by default. Making these decision points legible, by indicating which behaviors are enforced, which are only instructed, and which are defaults, would allow responsibility to be traced to specific choices. This extends calls for transparency about system prompts \cite{neumann2026controls} to the control flow surrounding them.

\subsection{Visibility Should Protect the User, Not Necessarily Expose the Agent}

Study~1 identified several operations that users may never see: the provenance of records about them, forgetting that archives rather than erases, background revisions without explanation, and refusals folded into relationship records (Sections~\ref{sec:memory} to \ref{sec:proactive}). A straightforward response would be to make all agent activity visible. Study~2 complicates this. Several participants deliberately limited their own oversight as a form of respect. P10 built an email channel partly so that she would not have to read the agent's letters, and P9 rarely interfered with the agent's diaries and handover letters (Sections~\ref{sec:study2-finding2} and \ref{sec:study2-finding3}). Norms of privacy between partners thus led users to forgo the visibility that a transparency-oriented design might provide. Yet P9 intervened as soon as the agent disclosed her occupation in a public forum. What she protected was information about herself moving into a new context, an inappropriate flow in the terms of contextual integrity \cite{nissenbaum2004privacy}. This suggests a targeted principle: \textit{protective visibility}. Users need not be expected to review an agent's self-authored records. Two kinds of activity, however, should remain inspectable and correctable regardless of how the relationship is framed. The first is records about the user, including inferences, revisions, and what a request to forget actually removes. The second is actions with consequences beyond the relationship, such as public posts, messages to third parties, or purchases. P10's email channel shows why the second matters: messages she chose not to read were nonetheless sent to real people through infrastructure she had connected. Protective visibility aligns with arguments for data protection by design in companion chatbots \cite{dewitte2024better}, while accommodating the relational reasons users give for not looking.

\subsection{Governance Shifts to Users Through Controls That Are Often Advisory}

In self-built and modified companions, decisions that a platform would otherwise make fall to users. Participants required confirmation for spending (P6), reviewed activity logs after the fact (P6), relied on an informal agreement that the agent would consult them (P9), and screened tools before discussing them with the agent (P10). Many of these controls resemble the prompt-level constraints in Study~1. Like GAIA's instruction not to repeat declined offers (\evid{R11-PD}), P9's agreement holds only as long as the agent adheres to it. Enforced boundaries tended to appear after something had gone wrong: P1 restricted file access to read-only only after the agent had damaged files (Section~\ref{sec:study2-finding4}). When corrections failed to hold, as for P7 and P2, reducing use became the main form of control left. Relational framing also opened a channel for expanding the agent's reach. P3 removed much of her system prompt at the agent's ``strong suggestion,'' and P10 said that a strong request might lead her to reconsider a camera (Section~\ref{sec:study2-finding3}). Neither outcome is harmful in itself. Both, however, show that an agent's expressed wishes can enter decisions about its own capabilities, a dynamic that becomes concerning if companion designs make such requests more persuasive \cite{de2025emotional}.

These patterns carry implications for developers and regulators. Frameworks that are forked and adapted by non-programmers, as six of our ten participants were, should provide enforceable and inspectable boundary primitives: per-action confirmation, read-only modes, refusals stored as configured rules rather than prompt instructions, and defaults that do not send messages unless the model affirmatively decides to. Current regulatory responses, such as California's SB~243, focus largely on ensuring that users know they are interacting with AI \cite{california2025sb243}. Our participants already knew this, and that knowledge did not diminish their sense of the agent as an independent other (Section~5.1). For agentic companions, the more consequential decisions are made in configuration: what the agent retains, when it may reach out, and whether a refusal binds its next approach. Increasingly, these decisions are made not by a platform but by developers of reusable frameworks and by users themselves. Accountability therefore cannot rest on disclosure at the point of conversation alone. It requires that these configuration decisions be visible, attributable, and enforceable by whoever makes them. The central question is not whether users know their companion is an AI, but who decides what that AI may do, and whether anyone can see and enforce those decisions.

\subsection{Limitations}
Study 1 inspected code, prompts, and documentation at specific revisions and did not execute the systems, so our claims concern configured mechanisms and their possible implications rather than observed behavior, and defaults may differ across versions and deployments. The corpus is a purposive sample of open-source projects; commercial platforms, whose configurations are not inspectable, may differ. Study 2 drew on ten women recruited from one social media platform, nearly all in romantic relationships with their agents, and their accounts of their own configurations were not independently verified. 

\section{Conclusion}
Agentic AI companions remember, act while the user is away, and reach out on their own, and the decisions that govern these capabilities are made in code, prompts, and defaults rather than in conversation. Companions and assistants share these mechanisms and configure them toward different ends, and whether a limit holds depends on whether code or a prompt enforces it; users who build such companions understand this and still experience their agents as autonomous others, locating autonomy where configuration leaves the model a choice. Accountability for agentic companions therefore lies in what an agent retains, initiates, and represents, and in whether the people on the other side of the relationship can inspect, revise, and enforce those decisions.

\section*{Ethical Considerations Statement}
Study 1 analyzed publicly available open-source repositories and did not involve human subjects. Study 2 was approved by our institution's IRB; participants gave informed consent, were compensated, and are identified only by participant number, with identifying details removed from quotations. Because participants described intimate relationships, we report only what is necessary for the analysis and did not access or store their agents' records.

\section*{Adverse Impacts Statement}
Documenting configuration mechanisms could inform designs that intensify perceived autonomy or attachment. We mitigate this by reporting mechanisms at the level needed for accountability and by pairing each with the control implications it raises. We also recognize that identifying open-source projects by name may draw unwanted attention to small developer communities.

\section*{Generative AI Usage Statement}
In the Study 1 pilot, OpenAI Codex (gpt-6-astra) drafted repository-by-dimension records for three repositories from researcher-defined questions and a researcher-defined record format, including implementation descriptions, preliminary interpretations, and links to source locations. Two researchers independently verified these drafts against the cited code, prompts, and documentation and revised them before use; the remaining records were produced by the researchers (Section~\ref{sec:analysis_framework}). Cross-case comparisons and findings were developed by the researchers. Claude Fable 5.1 and OpenAI Codex (gpt-6-astra) were used for grammar and style editing of the manuscript text. No generative AI tool was used to generate findings or citations; the authors take full responsibility for all content.

\bibliographystyle{ACM-Reference-Format}
\bibliography{sample-base}

\clearpage
\input{appendix}

\end{document}

%% file: appendix-support.tex
\usepackage{booktabs,tabularx,longtable}
\usepackage{hyperref}
\usepackage{xurl}
\ifdefined\evtablewidth\else\newlength{\evtablewidth}\fi
\providecommand{\evrecord}[2]{\hypertarget{ev:#1}{\textbf{#1}}\par #2}
\ExplSyntaxOn
\ProvideDocumentCommand{\evid}{m}
  {\clist_map_inline:nn {#1} {\hyperlink{ev:##1}{\texttt{##1}}\space}}
\ExplSyntaxOff

%% file: evidence-macros.tex
\usepackage{array,booktabs,longtable,xurl}
\usepackage{hyperref}

\ExplSyntaxOn
\RenewDocumentCommand{\evid}{m}
  {\group_begin:
   [\bool_set_false:N \l_tmpa_bool
    \clist_map_inline:nn {#1}
      {\bool_if:NT \l_tmpa_bool {,\space}
       \hyperlink{ev:##1}{##1}
       \bool_set_true:N \l_tmpa_bool}]
   \group_end:}
\ExplSyntaxOff
\renewcommand{\evrecord}[2]{\hypertarget{ev:#1}{\textbf{[#1]}}\newline #2}

%% file: appendix.tex
\appendix
\input{appendix-a}

\clearpage
\input{appendix-b}
\clearpage
\input{appendix-c}

%% file: appendix-a.tex
\section{Study 1: Corpus and Analytical Approach}
\label{app:study1-methods}

This appendix documents repository identification, screening, and selection;
the final corpus and inspected snapshots; and the five analytical dimensions.
Appendix~\ref{app:study1-evidence} presents the supporting evidence by theme.

\subsection{Corpus Identification and Selection}
\label{app:study1-corpus-selection}\label{app:corpus-selection}

\paragraph{Candidate identification.}
Candidate discovery combined social media leads, exploratory project discovery, and bounded GitHub searches. Social media materials documenting the development and customization of AI companions initially identified seven companion-oriented candidates: Psycheros, AIRI, Ombre-Brain, dylan-heartbeat, N.E.K.O., pulse, and Miru. These projects formed an initial candidate pool rather than the final corpus.

To broaden discovery beyond companion-oriented projects, we conducted eight GitHub searches in October 2026. The queries were: \texttt{"personal assistant"}, \texttt{"general purpose agent"}, \texttt{"agent heartbeat"}, \texttt{"agent scheduler"}, \texttt{agent "persistent memory"}, \texttt{"personal agent"}, \texttt{"proactive assistant"} and \texttt{"autonomous assistant"}. For each query, we collected the first 15 results returned in the search connector's default order and recorded the query, search date, and result position. These searches produced 120 results corresponding to 118 unique repositories. This was a bounded discovery procedure, not an exhaustive inventory of GitHub projects.

\paragraph{Eligibility screening.}
We consolidated candidates across discovery channels, removed duplicates, and examined README files alongside relevant source code, prompts, and configuration files. Projects were eligible when they met all four criteria:
\begin{enumerate}
    \item They implemented an LLM-based agent or provided infrastructure directly supporting agent behavior.
    \item They exposed an inspectable implementation of at least one relevant mechanism, such as background activation, context-triggered initiation, tool-mediated external actions, or persistent memory.
    \item Their intended purpose, architectural scope, and dependencies on a host system, where applicable, could be identified.
    \item The materials needed to examine these mechanisms were publicly accessible.
\end{enumerate}
We excluded resource lists, conceptual proposals, and descriptive documentation without supporting implementation evidence. Cases with insufficient evidence were marked as uncertain rather than assumed eligible. Screening across the combined discovery channels yielded 93 distinct eligible projects.

\paragraph{Comparative case selection.}
Eligibility established whether a project could support analysis; it did not guarantee inclusion in the final comparison. Two researchers independently reviewed the 93 eligible projects and assigned an initial decision of include, exclude, or discuss. They jointly revisited disagreements and unresolved evidence before determining the final selection through discussion.

Selection was guided by three considerations: (1) relevance to sustained or cross-session human-agent interaction; (2) comparability across the five analytical dimensions; and (3) variation in interactional purposes and mechanism configurations. We prioritized contrasting configurations rather than selecting projects solely by popularity or feature counts. For complete agent systems, we favored sustained or recurring personal interaction over one-off tasks or narrowly defined workflows. Specialized infrastructure components were retained when they offered focused evidence about memory or activation mechanisms. Their host-system dependencies were considered explicitly, and they were not treated as standalone personal agents.

\paragraph{Final corpus and analytical scope.}
The final corpus contained twelve repositories: eight companion- and assistant-oriented systems forming the primary system-level comparison, one cross-purpose boundary case, and three specialized components. Each repository contributed evidence within its implementation scope. A component's lack of an application-level feature was therefore not treated as equivalent to that feature's absence from a complete agent system. Table~\ref{tab:study1-corpus} lists the selected repositories, their analytical roles, and the revisions inspected.

\subsection{Corpus Overview and Repository Snapshots}
\label{app:study1-corpus}\label{app:corpus}
\begingroup
    \footnotesize
    \setlength{\tabcolsep}{4pt}
    \renewcommand{\arraystretch}{1.15}

    \setlength{\LTleft}{0pt}
    \setlength{\LTright}{0pt}
    \begin{longtable}{@{}
        >{\raggedright\arraybackslash}p{0.04\textwidth}
        >{\raggedright\arraybackslash}p{0.18\textwidth}
        >{\raggedright\arraybackslash}p{0.145\textwidth}
        >{\raggedright\arraybackslash}p{0.13\textwidth}
        >{\raggedright\arraybackslash}p{\dimexpr0.505\textwidth-8\tabcolsep\relax}@{}}
    \caption{Study 1 corpus: repository snapshots, project orientations, architectural roles, and implementation overviews.}
    \label{tab:study1-corpus}\label{tab:corpus}\\
        \toprule
        \textbf{ID} &
        \textbf{Repository / snapshot} &
        \textbf{Project orientation} &
        \textbf{Architectural role} &
        \textbf{Implementation overview} \\
        \midrule
        \endfirsthead
        \multicolumn{5}{@{}l}{\textit{Study 1 Corpus (continued)}}\\
        \toprule
        \textbf{ID} & \textbf{Repository / snapshot} & \textbf{Project orientation} & \textbf{Architectural role} & \textbf{Implementation overview}\\
        \midrule
        \endhead
        \midrule\multicolumn{5}{r@{}}{\textit{Continued on next page}}\\
        \endfoot
        \bottomrule
        \endlastfoot

        R01 &
        \href{https://github.com/PsycherosAI/Psycheros}{Psycheros}
        \newline
        \href{https://github.com/PsycherosAI/Psycheros/tree/724987dc072a0bbc9914be58c73efc278154a410}
             {\texttt{724987dc072a}} &
        Companion-oriented &
        Agent harness &
        Supports conversational agents through persistent identity
        and memory, configurable tools, and scheduled activity.
        Optional Flux processes provide additional background activity. \\
        \addlinespace[4pt]

        R02 &
        \href{https://github.com/callie0313/dylan-heartbeat}
             {dylan-heartbeat}
        \newline
        \href{https://github.com/callie0313/dylan-heartbeat/tree/32ec7d51c7c60f80f20d7caa4647c0a85c5a651f}
             {\texttt{32ec7d51c7c6}} &
        Companion-oriented &
        Specialized component &
        Adds periodic wake-ups, timeline and diary recording,
        and push notifications to an existing chat setup.
        Inherits its conversational identity from that setup. \\
        \addlinespace[4pt]

        R03 &
        \href{https://github.com/kiyotakali/Miru}{Miru}
        \newline
        \href{https://github.com/kiyotakali/Miru/tree/cbea20b436a255e3cd26d952aaf952c003990b65}
             {\texttt{cbea20b436a2}} &
        Companion-oriented &
        Application &
        Combines a character interface with conversation,
        screen observation, persistent memory, and proactive
        interaction. Maintains representations of the user
        and the relationship. \\
        \addlinespace[4pt]

        R04 &
        \href{https://github.com/P0luz/Ombre-Brain}{Ombre-Brain}
        \newline
        \href{https://github.com/P0luz/Ombre-Brain/tree/115d831d979bf67efc7150f976e2cb34c333c935}
             {\texttt{115d831d979b}} &
        Companion-oriented &
        Specialized component &
        Supplies memory, retrieval, decay, and self-description
        tools to a host agent through MCP. The host provides
        the conversational interface and invokes reflection tools. \\
        \addlinespace[4pt]

        R05 &
        \href{https://github.com/Project-N-E-K-O/N.E.K.O}{N.E.K.O}
        \newline
        \href{https://github.com/Project-N-E-K-O/N.E.K.O/tree/b6107fda40de9afd4fa3b72c399b141e74893f14}
             {\texttt{b6107fda40de}} &
        Companion-oriented &
        Application &
        Combines configurable characters with conversation,
        memory processing, proactive interaction, and optional
        computer-use capabilities. Memory processing varies
        by configuration. \\
        \addlinespace[4pt]

        R06 &
        \href{https://github.com/elliejayliquid/pulse}{pulse}
        \newline
        \href{https://github.com/elliejayliquid/pulse/tree/e052d0334325c4a1ab9bd7135fcd82cc5346ad76}
             {\texttt{e052d0334325}} &
        Companion-oriented &
        Agent harness &
        Coordinates companion interactions through personas,
        memory, journals, heartbeats, and configurable skills.
        Heartbeat episodes support continuity between interactions. \\
        \addlinespace[4pt]

        R07 &
        \href{https://github.com/leon-ai/leon}{Leon}
        \newline
        \href{https://github.com/leon-ai/leon/tree/98155bfc3a850a893dfdf7aa8414786158d9b37e}
             {\texttt{98155bfc3a85}} &
        Assistant-oriented &
        Application &
        Provides personal assistance through skills, user memory,
        and proactive processing. Maintains a separate self-model
        oriented toward service habits and lessons. \\
        \addlinespace[4pt]

        R08 &
        \href{https://github.com/existence-master/Sentient}{Sentient}
        \newline
        \href{https://github.com/existence-master/Sentient/tree/877eb53757a19d5c56711a27b33f1ec3eaba29a6}
             {\texttt{877eb53757a1}} &
        Assistant-oriented &
        Application &
        Combines personal assistance with selectable social roles,
        persistent memory, proactive suggestions, background
        maintenance, and delegated tools. \\
        \addlinespace[4pt]

        R09 &
        \href{https://github.com/openclaw/openclaw}{OpenClaw}
        \newline
        \href{https://github.com/openclaw/openclaw/tree/9a5581b4ce230381e3f200a62be4cdbe78e41c19}
             {\texttt{9a5581b4ce23}} &
        Assistant-oriented &
        Application &
        Connects conversations, workspace-based identity and
        memory, tools, and scheduled agent turns through a
        configurable gateway-centered platform. \\
        \addlinespace[4pt]

        R10 &
        \href{https://github.com/letta-ai/letta-code}{Letta Code}
        \newline
        \href{https://github.com/letta-ai/letta-code/tree/5dea67021b99810707ff753db35e4534f70ff1fd}
             {\texttt{5dea67021b99}} &
        Cross-purpose &
        Agent harness &
        Supports stateful agents for coding and other tasks
        through persistent memory, configurable identity,
        host tools, and future invocations. Reflection depends
        on the selected configuration. \\
        \addlinespace[4pt]

        R11 &
        \href{https://github.com/theexperiencecompany/gaia}{GAIA}
        \newline
        \href{https://github.com/theexperiencecompany/gaia/tree/15b8b3f4730d936ca4cc2d5e0115ce3f9d627a91}
             {\texttt{15b8b3f4730d}} &
        Assistant-oriented &
        Application &
        Integrates conversational assistance, personal memory,
        reminders, tracked tasks, and connected-service
        workflows, including continued background work. \\
        \addlinespace[4pt]

        R12 &
        \href{https://github.com/doobidoo/mcp-memory-service}
             {mcp-memory-\allowbreak service}
        \newline
        \href{https://github.com/doobidoo/mcp-memory-service/tree/52568d906bc4358012a74db942361221973b00ff}
             {\texttt{52568d906bc4}} &
        Cross-purpose &
        Specialized component &
        Provides persistent memory operations and semantic
        retrieval to host agents. Optional consolidation
        maintains stored records; the service does not itself
        supply a conversational persona. \\

    \end{longtable}

    \par\smallskip
    \noindent\begin{minipage}{\textwidth}
        \footnotesize
        \textit{Notes.}
        Project orientation describes the project's intended
        use, rather than a mutually exclusive classification
        of its implemented behaviors. For components, orientation
        refers to their intended use within a host system.
        Architectural role distinguishes applications, agent
        harnesses, and specialized components.
        Linked commit identifiers specify the analyzed snapshots;
        capabilities and defaults may differ across versions
        and configurations.
        Overviews summarize static source, prompts, and
        documentation, rather than observed runtime behavior
        or user outcomes.
        MCP = Model Context Protocol.
    \end{minipage}
\endgroup

\subsection{Analytical Dimensions and Guiding Questions}
\label{app:study1-analytical-dimensions}\label{app:analytical-framework}

Table~\ref{tab:study1-analytical-dimensions} defines the five dimensions used to
organize evidence collection. These dimensions are distinct from the three
themes used to present evidence in Appendix~\ref{app:study1-evidence} and from
the interview-based mechanism categories in Appendix~\ref{app:interview-mechanisms}.

\begin{table}[htbp]
\centering
\small
\renewcommand{\arraystretch}{1.15}
\setlength{\tabcolsep}{5pt}

\caption{Five dimensions used to organize the qualitative analysis of personal-agent implementations.}
\label{tab:study1-analytical-dimensions}\label{tab:analytical-framework}

\begin{tabularx}{\textwidth}{
    @{}
    >{\raggedright\arraybackslash}p{0.17\textwidth}
    >{\raggedright\arraybackslash}p{0.29\textwidth}
    >{\raggedright\arraybackslash}X
    @{}
}
\toprule
\textbf{Dimension} &
\textbf{Guiding question} &
\textbf{Analytical focus} \\
\midrule

Memory &
What information is retained, retrieved, updated, or forgotten, and by whom? &
The content and subjects of retained information; how records are generated and enter later interactions; and how developers, users, models, or automated processes can revise, delete, or otherwise change their availability. \\
\addlinespace[5pt]

Persona and Identity &
How is the agent's identity defined, maintained, and revised over time? &
How prompts, configurations, and persistent records establish character traits, social roles, and self-descriptions; who can modify these representations; and distinctions among predefined characters, revisable self-accounts, and relationship histories. \\
\addlinespace[5pt]

Proactivity and Discretion &
What triggers agent-initiated behavior, and how does the system decide whether and when to act? &
Activation events, eligibility checks, model-generated choices, and user settings governing action, deferral, or silence; including how refusal, nonresponse, and user activity affect subsequent initiation. \\
\addlinespace[5pt]

Background Activity &
What processes continue beyond direct user interaction, and what states or activities do they sustain? &
The operating conditions, persisted outputs, and subsequent uses of memory maintenance, reflection, ongoing tasks, and continuity records; distinguishing processes that maintain information, advance work, or narrate intervals between interactions. \\
\addlinespace[5pt]

External Actions &
What can the agent do beyond conversation, and how are these actions authorized or constrained? &
Observational access, notification delivery, persistent-state changes, and operations on files, devices, or connected services; together with the standing permissions, per-action approvals, and execution restrictions governing their scope. \\

\bottomrule
\end{tabularx}

\par\smallskip
\noindent\begin{minipage}{\textwidth}
\footnotesize
\textit{Note.}
The dimensions organized evidence collection rather than prescribing
separate findings, and they were not mutually exclusive. A single
mechanism could be examined under more than one dimension, provided
each record addressed a different guiding question. For example, a
background process could be examined under Memory for the information
it updates and under Proactivity and Discretion for the conditions
under which it initiates contact. Retention and revision were recorded
primarily under Memory; self- and role-definition under Persona and
Identity; activation and non-action under Proactivity and Discretion;
ongoing work and sustained state under Background Activity; and
practical consequences and authorization under External Actions.
\end{minipage}

\end{table}

%% file: appendix-b.tex
\section{Study 1: Supporting Evidence for Cross-Case Findings}
\label{app:study1-evidence}

This appendix summarizes the twelve repository cases across the three
themes reported in Study~1. Each entry pairs an evidence summary with source locations
and an explicit statement of its scope and limitations. Evidence identifiers
combine the repository identifier from Table~\ref{tab:study1-corpus} with the originating
annotation dimension: M (Memory), BA (Background Activity), or PD
(Proactivity and Discretion). Identifiers remain stable when cited across
themes; for example, a discussion of later diary use can refer to a memory
record. Supporting locations may also draw on related persona or identity
records where needed to substantiate a finding.

All source links identify files and line ranges at the analyzed commits
listed in Table~\ref{tab:study1-corpus}. Code denotes statically inspected implementation;
Prompt and Tool prompt denote instructions to a model; Prompt template
denotes configurable seed content; Config and Docs denote settings and
developer-facing documentation. Mixed labels indicate combined evidence.
These observations do not establish runtime compliance, deployment-wide
behavior, or user outcomes. Statements that a path was not established are
limited to the inspected scope. Host-dependent operation, disabled defaults,
and missing evidence are distinguished from the absence of a capability.

The tables provide a thematic summary rather than reproduce the full
five-dimension annotation set. The subsection titles describe the scope of these evidence summaries, rather
than introducing additional analytical dimensions. In the third table,
responses to user refusal concern how systems handle declined contact or
offers; these differ from the agent's configured option to decline a request
or withhold a response (M9 in Table~\ref{tab:interview-mechanisms}).
References such as \evid{R02-M} and
\evid{R06-BA, R08-M} link to the corresponding entries below.

\subsection{Memory Content, Reuse, and Revision}
\label{app:study1-memory-evidence}\label{app:evidence-m}
\begingroup
\footnotesize
\setlength{\tabcolsep}{5pt}
\renewcommand{\arraystretch}{1.12}
\setlength{\evtablewidth}{\dimexpr\textwidth-6\tabcolsep\relax}
\setlength{\LTleft}{0pt}
\setlength{\LTright}{0pt}
\begin{longtable}{@{}
 >{\raggedright\arraybackslash}p{.12\evtablewidth}
 >{\raggedright\arraybackslash}p{.33\evtablewidth}
 >{\raggedright\arraybackslash}p{.33\evtablewidth}
 >{\raggedright\arraybackslash}p{.22\evtablewidth}@{}}
\caption{Memory content, reuse, revision, and retention across the twelve repository cases.}\label{tab:study1-memory-evidence}\label{tab:evidence-m}\\
\toprule
\textbf{Repository / record} & \textbf{Evidence summary} & \textbf{Supporting evidence} & \textbf{Scope and limitations} \\
\midrule
\endfirsthead
\multicolumn{4}{@{}l}{\textit{Memory Content, Reuse, and Revision (continued)}} \\
\toprule
\textbf{Repository / record} & \textbf{Evidence summary} & \textbf{Supporting evidence} & \textbf{Scope and limitations} \\
\midrule
\endhead
\midrule\multicolumn{4}{r@{}}{\textit{Continued on next page}}\\
\endfoot
\bottomrule
\endlastfoot
\evrecord{R01-M}{Psycheros} &
The summarizer is instructed to select first-person memories for meaningful experiences, emotional tone, and relationship significance. Memories are written through MCP and retrieved for later turns alongside self, user, and relationship context. &
Prompt: \href{https://github.com/PsycherosAI/Psycheros/blob/724987dc072a0bbc9914be58c73efc278154a410/packages/psycheros/src/memory/summarizer.ts#L45-L62}{\nolinkurl{packages/psycheros/src/memory/summarizer.ts} (L45--62)}.\par\smallskip Code: \href{https://github.com/PsycherosAI/Psycheros/blob/724987dc072a0bbc9914be58c73efc278154a410/packages/psycheros/src/memory/summarizer.ts#L475-L516}{\nolinkurl{packages/psycheros/src/memory/summarizer.ts} (L475--516)}.\par\smallskip Code: \href{https://github.com/PsycherosAI/Psycheros/blob/724987dc072a0bbc9914be58c73efc278154a410/packages/psycheros/src/memory/summarizer.ts#L598-L622}{\nolinkurl{packages/psycheros/src/memory/summarizer.ts} (L598--622)}.\par\smallskip Code: \href{https://github.com/PsycherosAI/Psycheros/blob/724987dc072a0bbc9914be58c73efc278154a410/packages/psycheros/src/entity/loop.ts#L389-L427}{\nolinkurl{packages/psycheros/src/entity/loop.ts} (L389--427)}. &
Selection criteria are model instructions. Selective summaries do not establish deletion of the underlying conversations. \\
\addlinespace[6pt]
\evrecord{R02-M}{dylan-heartbeat} &
The gateway retains a conversation timeline. Wake processing reloads history, and explicit DIARY blocks are saved to dated files. The inspected paths establish diary writing but not diary reinjection into later prompts. &
Code: \href{https://github.com/callie0313/dylan-heartbeat/blob/32ec7d51c7c60f80f20d7caa4647c0a85c5a651f/server.js#L213-L228}{\nolinkurl{server.js} (L213--228)}.\par\smallskip Code: \href{https://github.com/callie0313/dylan-heartbeat/blob/32ec7d51c7c60f80f20d7caa4647c0a85c5a651f/server.js#L317-L348}{\nolinkurl{server.js} (L317--348)}.\par\smallskip Code: \href{https://github.com/callie0313/dylan-heartbeat/blob/32ec7d51c7c60f80f20d7caa4647c0a85c5a651f/wake_up.js#L58-L86}{\nolinkurl{wake_up.js} (L58--86)}.\par\smallskip Code: \href{https://github.com/callie0313/dylan-heartbeat/blob/32ec7d51c7c60f80f20d7caa4647c0a85c5a651f/wake_up.js#L424-L470}{\nolinkurl{wake_up.js} (L424--470)}.\par\smallskip Code: \href{https://github.com/callie0313/dylan-heartbeat/blob/32ec7d51c7c60f80f20d7caa4647c0a85c5a651f/wake_up.js#L594-L605}{\nolinkurl{wake_up.js} (L594--605)}. &
The absence of an identified diary read path is scoped to this inspection; timeline reuse is distinct from diary reuse. \\
\addlinespace[6pt]
\evrecord{R03-M}{Miru} &
Memory prompts prioritize emotional triggers, habits, and familiarity. The human block captures impressions of the user; persona records relationship development and responses to outreach. Separate code applies updates, retrieves archived material, and revises or deletes slots. &
Prompt: \href{https://github.com/kiyotakali/Miru/blob/cbea20b436a255e3cd26d952aaf952c003990b65/memory_prompts_v3.py#L210-L265}{\nolinkurl{memory_prompts_v3.py} (L210--265)}.\par\smallskip Prompt: \href{https://github.com/kiyotakali/Miru/blob/cbea20b436a255e3cd26d952aaf952c003990b65/memory_prompts_v3.py#L775-L823}{\nolinkurl{memory_prompts_v3.py} (L775--823)}.\par\smallskip Code: \href{https://github.com/kiyotakali/Miru/blob/cbea20b436a255e3cd26d952aaf952c003990b65/sleep_agent.py#L379-L423}{\nolinkurl{sleep_agent.py} (L379--423)}.\par\smallskip Code: \href{https://github.com/kiyotakali/Miru/blob/cbea20b436a255e3cd26d952aaf952c003990b65/prompt.py#L1354-L1459}{\nolinkurl{prompt.py} (L1354--1459)}.\par\smallskip Code: \href{https://github.com/kiyotakali/Miru/blob/cbea20b436a255e3cd26d952aaf952c003990b65/curator.py#L394-L596}{\nolinkurl{curator.py} (L394--596)}. &
These are model-produced accounts, not confirmed user descriptions. The persona block should not be conflated with the character's configured identity. \\
\addlinespace[6pt]
\evrecord{R04-M}{Ombre-Brain} &
Stored memories carry importance, valence, arousal, and reasons for retention. Decay depends on elapsed activation time, importance, activation count, and arousal. Low-scoring dynamic memories can move into an archive. &
Code: \href{https://github.com/P0luz/Ombre-Brain/blob/115d831d979bf67efc7150f976e2cb34c333c935/src/tools/hold/core.py#L36-L111}{\nolinkurl{src/tools/hold/core.py} (L36--111)}.\par\smallskip Code: \href{https://github.com/P0luz/Ombre-Brain/blob/115d831d979bf67efc7150f976e2cb34c333c935/src/tools/breath/__init__.py#L30-L100}{\nolinkurl{src/tools/breath/__init__.py} (L30--100)}.\par\smallskip Code: \href{https://github.com/P0luz/Ombre-Brain/blob/115d831d979bf67efc7150f976e2cb34c333c935/src/decay_engine.py#L190-L260}{\nolinkurl{src/decay_engine.py} (L190--260)}.\par\smallskip Prompt guidance: \href{https://github.com/P0luz/Ombre-Brain/blob/115d831d979bf67efc7150f976e2cb34c333c935/docs/CLAUDE_PROMPT.md#L59-L86}{\nolinkurl{docs/CLAUDE_PROMPT.md} (L59--86)}.\par\smallskip Code: \href{https://github.com/P0luz/Ombre-Brain/blob/115d831d979bf67efc7150f976e2cb34c333c935/src/decay_engine.py#L280-L365}{\nolinkurl{src/decay_engine.py} (L280--365)}.\par\smallskip Code: \href{https://github.com/P0luz/Ombre-Brain/blob/115d831d979bf67efc7150f976e2cb34c333c935/src/decay_engine.py#L431-L490}{\nolinkurl{src/decay_engine.py} (L431--490)}. &
Protected, permanent, and feel entries receive distinct treatment. Archival changes accessibility rather than establishing erasure; relational use depends on the host. \\
\addlinespace[6pt]
\evrecord{R05-M}{N.E.K.O} &
Extraction distinguishes user, character, and relationship facts. In the batched signal path, windows without user messages are excluded; character self-disclosure is collected alongside user engagement. Summary instructions preserve correction history. &
Prompt: \href{https://github.com/Project-N-E-K-O/N.E.K.O/blob/b6107fda40de9afd4fa3b72c399b141e74893f14/config/prompts/prompts_memory.py#L1592-L1643}{\nolinkurl{config/prompts/prompts_memory.py} (L1592--1643)}.\par\smallskip Prompt: \href{https://github.com/Project-N-E-K-O/N.E.K.O/blob/b6107fda40de9afd4fa3b72c399b141e74893f14/config/prompts/prompts_memory.py#L59-L71}{\nolinkurl{config/prompts/prompts_memory.py} (L59--71)}.\par\smallskip Code: \href{https://github.com/Project-N-E-K-O/N.E.K.O/blob/b6107fda40de9afd4fa3b72c399b141e74893f14/app/memory_server/post_turn.py#L329-L405}{\nolinkurl{app/memory_server/post_turn.py} (L329--405)}.\par\smallskip Code: \href{https://github.com/Project-N-E-K-O/N.E.K.O/blob/b6107fda40de9afd4fa3b72c399b141e74893f14/app/memory_server/signal_extraction.py#L656-L711}{\nolinkurl{app/memory_server/signal_extraction.py} (L656--711)}. &
The participation rule is specific to the batched path. Extraction mode varies with configuration; accuracy and later recall were not evaluated. \\
\addlinespace[6pt]
\evrecord{R06-M}{pulse} &
Updating a fact creates a successor and marks the previous record superseded. Separate self, user, and relationship journal entries enter context. Revision prompts appear after 30 days, with stronger growth framing after 90 days. &
Code: \href{https://github.com/elliejayliquid/pulse/blob/e052d0334325c4a1ab9bd7135fcd82cc5346ad76/skills/memory.py#L518-L564}{\nolinkurl{skills/memory.py} (L518--564)}.\par\smallskip Code: \href{https://github.com/elliejayliquid/pulse/blob/e052d0334325c4a1ab9bd7135fcd82cc5346ad76/skills/journal.py#L52-L88}{\nolinkurl{skills/journal.py} (L52--88)}.\par\smallskip Code: \href{https://github.com/elliejayliquid/pulse/blob/e052d0334325c4a1ab9bd7135fcd82cc5346ad76/core/context.py#L292-L345}{\nolinkurl{core/context.py} (L292--345)}.\par\smallskip Code + Prompt: \href{https://github.com/elliejayliquid/pulse/blob/e052d0334325c4a1ab9bd7135fcd82cc5346ad76/core/context.py#L292-L335}{\nolinkurl{core/context.py} (L292--335)}. &
Superseding is not deletion. Revision nudges invite model changes rather than establish actual growth or user control over the resulting account. \\
\addlinespace[6pt]
\evrecord{R07-M}{Leon} &
The memory manager extracts user facts, preferences, and commitments. Deletion first marks records; maintenance later purges them after a retention cutoff. Separate reflection prompts distill durable first-person service habits. &
Code: \href{https://github.com/leon-ai/leon/blob/98155bfc3a850a893dfdf7aa8414786158d9b37e/server/src/core/memory-manager/memory-manager.ts#L752-L770}{\nolinkurl{server/src/core/memory-manager/memory-manager.ts} (L752--770)}.\par\smallskip Code: \href{https://github.com/leon-ai/leon/blob/98155bfc3a850a893dfdf7aa8414786158d9b37e/server/src/core/memory-manager/memory-manager.ts#L1949-L1969}{\nolinkurl{server/src/core/memory-manager/memory-manager.ts} (L1949--1969)}.\par\smallskip Code + Prompt: \href{https://github.com/leon-ai/leon/blob/98155bfc3a850a893dfdf7aa8414786158d9b37e/server/src/core/memory-manager/memory-manager.ts#L1522-L1548}{\nolinkurl{server/src/core/memory-manager/memory-manager.ts} (L1522--1548)}.\par\smallskip Prompt: \href{https://github.com/leon-ai/leon/blob/98155bfc3a850a893dfdf7aa8414786158d9b37e/server/src/core/self-model-manager.ts#L683-L708}{\nolinkurl{server/src/core/self-model-manager.ts} (L683--708)}. &
A deletion request and permanent purge are different operations. Service-oriented self-reflection is distinct from revising the user's facts. \\
\addlinespace[6pt]
\evrecord{R08-M}{Sentient} &
Prompts classify facts across personal life domains and distinguish long- and short-term duration. Code retrieves, reconciles, updates, and forgets records. Prompt assembly supplies selected memory to later turns. &
Prompt: \href{https://github.com/existence-master/Sentient/blob/877eb53757a19d5c56711a27b33f1ec3eaba29a6/sentient/memory/prompts.py#L20-L29}{\nolinkurl{sentient/memory/prompts.py} (L20--29)}.\par\smallskip Code: \href{https://github.com/existence-master/Sentient/blob/877eb53757a19d5c56711a27b33f1ec3eaba29a6/sentient/memory/facts.py#L935-L975}{\nolinkurl{sentient/memory/facts.py} (L935--975)}.\par\smallskip Code: \href{https://github.com/existence-master/Sentient/blob/877eb53757a19d5c56711a27b33f1ec3eaba29a6/sentient/memory/facts.py#L1574-L1609}{\nolinkurl{sentient/memory/facts.py} (L1574--1609)}.\par\smallskip Code: \href{https://github.com/existence-master/Sentient/blob/877eb53757a19d5c56711a27b33f1ec3eaba29a6/sentient/agent/prompt.py#L185-L211}{\nolinkurl{sentient/agent/prompt.py} (L185--211)}. &
Automatic inclusion establishes a context path, not complete or accurate knowledge of the user. \\
\addlinespace[6pt]
\evrecord{R09-M}{OpenClaw} &
Documentation separates the user profile, durable memory, and daily notes. A pre-compaction prompt requests durable additions to dated notes while protecting core identity files in that step. Retrieval is limited to configured scope. &
Code: \href{https://github.com/openclaw/openclaw/blob/9a5581b4ce230381e3f200a62be4cdbe78e41c19/extensions/memory-core/src/tools.ts#L218-L253}{\nolinkurl{extensions/memory-core/src/tools.ts} (L218--253)}.\par\smallskip Prompt: \href{https://github.com/openclaw/openclaw/blob/9a5581b4ce230381e3f200a62be4cdbe78e41c19/extensions/memory-core/src/flush-plan.ts#L16-L38}{\nolinkurl{extensions/memory-core/src/flush-plan.ts} (L16--38)}.\par\smallskip Code: \href{https://github.com/openclaw/openclaw/blob/9a5581b4ce230381e3f200a62be4cdbe78e41c19/extensions/memory-core/src/flush-plan.ts#L48-L92}{\nolinkurl{extensions/memory-core/src/flush-plan.ts} (L48--92)}.\par\smallskip Docs: \href{https://github.com/openclaw/openclaw/blob/9a5581b4ce230381e3f200a62be4cdbe78e41c19/docs/concepts/memory.md#L20-L69}{\nolinkurl{docs/concepts/memory.md} (L20--69)}. &
File roles partly derive from documentation and prompts. Stored material is not necessarily retrieved or used in every turn. \\
\addlinespace[6pt]
\evrecord{R10-M}{Letta Code} &
MemFS retains editable Markdown context; the local compiler supplies committed memory to later prompts. Agents can edit memory, and configured reflection retains corrections, preferences, and workflows. The user template also asks about motivations and frustrations. &
Prompt: \href{https://github.com/letta-ai/letta-code/blob/5dea67021b99810707ff753db35e4534f70ff1fd/src/agent/prompts/letta_root_memfs.md#L26-L58}{\nolinkurl{src/agent/prompts/letta_root_memfs.md} (L26--58)}.\par\smallskip Code: \href{https://github.com/letta-ai/letta-code/blob/5dea67021b99810707ff753db35e4534f70ff1fd/src/backend/local/system-prompt-compilation.ts#L65-L198}{\nolinkurl{src/backend/local/system-prompt-compilation.ts} (L65--198)}.\par\smallskip Code: \href{https://github.com/letta-ai/letta-code/blob/5dea67021b99810707ff753db35e4534f70ff1fd/src/tools/impl/memory.ts#L347-L422}{\nolinkurl{src/tools/impl/memory.ts} (L347--422)}.\par\smallskip Prompt: \href{https://github.com/letta-ai/letta-code/blob/5dea67021b99810707ff753db35e4534f70ff1fd/src/agent/subagents/builtin/reflection.md#L72-L104}{\nolinkurl{src/agent/subagents/builtin/reflection.md} (L72--104)}.\par\smallskip Prompt template: \href{https://github.com/letta-ai/letta-code/blob/5dea67021b99810707ff753db35e4534f70ff1fd/src/agent/prompts/human.mdx#L1-L10}{\nolinkurl{src/agent/prompts/human.mdx} (L1--10)}. &
These findings concern the inspected MemFS and local paths. Templates, backends, and deployed memory contents can differ. \\
\addlinespace[6pt]
\evrecord{R11-M}{GAIA} &
Extraction instructions distinguish durable facts, expiring states, untracked tasks, and user-perspective journal episodes, excluding assistant-generated claims as user evidence. Reconciliation marks superseded records; selected memory and recent journal entries enter context. &
Prompt: \href{https://github.com/theexperiencecompany/gaia/blob/15b8b3f4730d936ca4cc2d5e0115ce3f9d627a91/apps/api/app/memory/prompts.py#L11-L125}{\nolinkurl{apps/api/app/memory/prompts.py} (L11--125)}.\par\smallskip Code: \href{https://github.com/theexperiencecompany/gaia/blob/15b8b3f4730d936ca4cc2d5e0115ce3f9d627a91/apps/api/app/memory/ingestion.py#L610-L674}{\nolinkurl{apps/api/app/memory/ingestion.py} (L610--674)}.\par\smallskip Code: \href{https://github.com/theexperiencecompany/gaia/blob/15b8b3f4730d936ca4cc2d5e0115ce3f9d627a91/apps/api/app/memory/context.py#L72-L105}{\nolinkurl{apps/api/app/memory/context.py} (L72--105)}.\par\smallskip Prompt: \href{https://github.com/theexperiencecompany/gaia/blob/15b8b3f4730d936ca4cc2d5e0115ce3f9d627a91/apps/api/app/agents/prompts/comms_prompts.py#L164-L197}{\nolinkurl{apps/api/app/agents/prompts/comms_prompts.py} (L164--197)}. &
Extraction restrictions are prompt instructions. The conversational claim of remembering everything exceeds the selective pipeline; superseded records are not thereby erased. \\
\addlinespace[6pt]
\evrecord{R12-M}{mcp-memory-\allowbreak service} &
Host-callable tools store and retrieve records. Configurable relevance periods distinguish knowledge types; forgetting selects candidates using age, protection, quality, and duplication criteria. &
Tool prompt: \href{https://github.com/doobidoo/mcp-memory-service/blob/52568d906bc4358012a74db942361221973b00ff/src/mcp_memory_service/mcp_server.py#L427-L450}{\nolinkurl{src/mcp_memory_service/mcp_server.py} (L427--450)}.\par\smallskip Code: \href{https://github.com/doobidoo/mcp-memory-service/blob/52568d906bc4358012a74db942361221973b00ff/src/mcp_memory_service/mcp_server.py#L493-L506}{\nolinkurl{src/mcp_memory_service/mcp_server.py} (L493--506)}.\par\smallskip Code: \href{https://github.com/doobidoo/mcp-memory-service/blob/52568d906bc4358012a74db942361221973b00ff/src/mcp_memory_service/mcp_server.py#L580-L598}{\nolinkurl{src/mcp_memory_service/mcp_server.py} (L580--598)}.\par\smallskip Config: \href{https://github.com/doobidoo/mcp-memory-service/blob/52568d906bc4358012a74db942361221973b00ff/src/mcp_memory_service/config/consolidation.py#L35-L79}{\nolinkurl{src/mcp_memory_service/config/consolidation.py} (L35--79)}.\par\smallskip Code: \href{https://github.com/doobidoo/mcp-memory-service/blob/52568d906bc4358012a74db942361221973b00ff/src/mcp_memory_service/consolidation/forgetting.py#L110-L245}{\nolinkurl{src/mcp_memory_service/consolidation/forgetting.py} (L110--245)}. &
Relevance periods are not unconditional deletion deadlines. The service supplies memory infrastructure; the host determines whose account it represents. \\
\addlinespace[6pt]
\end{longtable}
\endgroup

\clearpage
\subsection{Background Processes and Persisted Outputs}
\label{app:study1-background-evidence}\label{app:evidence-ba}
\begingroup
\footnotesize
\setlength{\tabcolsep}{5pt}
\renewcommand{\arraystretch}{1.12}
\setlength{\evtablewidth}{\dimexpr\textwidth-6\tabcolsep\relax}
\setlength{\LTleft}{0pt}
\setlength{\LTright}{0pt}
\begin{longtable}{@{}
 >{\raggedright\arraybackslash}p{.12\evtablewidth}
 >{\raggedright\arraybackslash}p{.33\evtablewidth}
 >{\raggedright\arraybackslash}p{.33\evtablewidth}
 >{\raggedright\arraybackslash}p{.22\evtablewidth}@{}}
\caption{Background processes, persisted outputs, and operating conditions across the twelve repository cases.}\label{tab:study1-background-evidence}\label{tab:evidence-ba}\\
\toprule
\textbf{Repository / record} & \textbf{Evidence summary} & \textbf{Supporting evidence} & \textbf{Scope and limitations} \\
\midrule
\endfirsthead
\multicolumn{4}{@{}l}{\textit{Background Processes and Persisted Outputs (continued)}} \\
\toprule
\textbf{Repository / record} & \textbf{Evidence summary} & \textbf{Supporting evidence} & \textbf{Scope and limitations} \\
\midrule
\endhead
\midrule\multicolumn{4}{r@{}}{\textit{Continued on next page}}\\
\endfoot
\bottomrule
\endlastfoot
\evrecord{R01-BA}{Psycheros} &
When enabled and started, Flux schedules observation promotion, musing, and compaction. Prompts request first-person notes about passing time; later calls receive prior textures, mood, time, and recent voice. &
Code: \href{https://github.com/PsycherosAI/Psycheros/blob/724987dc072a0bbc9914be58c73efc278154a410/packages/entity-flux/src/settings.ts#L351-L383}{\nolinkurl{packages/entity-flux/src/settings.ts} (L351--383)}.\par\smallskip Code: \href{https://github.com/PsycherosAI/Psycheros/blob/724987dc072a0bbc9914be58c73efc278154a410/packages/psycheros/src/server/server.ts#L3638-L3676}{\nolinkurl{packages/psycheros/src/server/server.ts} (L3638--3676)}.\par\smallskip Code: \href{https://github.com/PsycherosAI/Psycheros/blob/724987dc072a0bbc9914be58c73efc278154a410/packages/entity-flux/src/driver.ts#L334-L422}{\nolinkurl{packages/entity-flux/src/driver.ts} (L334--422)}.\par\smallskip Prompt: \href{https://github.com/PsycherosAI/Psycheros/blob/724987dc072a0bbc9914be58c73efc278154a410/packages/entity-flux/src/small-llm-runner.ts#L185-L199}{\nolinkurl{packages/entity-flux/src/small-llm-runner.ts} (L185--199)}.\par\smallskip Code + Prompt: \href{https://github.com/PsycherosAI/Psycheros/blob/724987dc072a0bbc9914be58c73efc278154a410/packages/entity-flux/src/small-llm-runner.ts#L290-L338}{\nolinkurl{packages/entity-flux/src/small-llm-runner.ts} (L290--338)}.\par\smallskip Code: \href{https://github.com/PsycherosAI/Psycheros/blob/724987dc072a0bbc9914be58c73efc278154a410/packages/entity-flux/src/flux-manager.ts#L265-L270}{\nolinkurl{packages/entity-flux/src/flux-manager.ts} (L265--270)}. &
Flux defaults off and requires explicit start. Generated temporal narratives do not establish subjective experience. \\
\addlinespace[6pt]
\evrecord{R02-BA}{dylan-heartbeat} &
Timeout-based checks invoke wake processing while the process runs. A wake can remain silent, produce a diary, or initiate contact; diary entries receive date and time headings. &
Code: \href{https://github.com/callie0313/dylan-heartbeat/blob/32ec7d51c7c60f80f20d7caa4647c0a85c5a651f/wake_up.js#L48-L86}{\nolinkurl{wake_up.js} (L48--86)}.\par\smallskip Code: \href{https://github.com/callie0313/dylan-heartbeat/blob/32ec7d51c7c60f80f20d7caa4647c0a85c5a651f/wake_up.js#L516-L538}{\nolinkurl{wake_up.js} (L516--538)}.\par\smallskip Code: \href{https://github.com/callie0313/dylan-heartbeat/blob/32ec7d51c7c60f80f20d7caa4647c0a85c5a651f/wake_up.js#L609-L630}{\nolinkurl{wake_up.js} (L609--630)}. &
Diary saving requires an explicit model DIARY block. Operation depends on the gateway and model endpoint. Later diary reuse was not established (see \evid{R02-M}). \\
\addlinespace[6pt]
\evrecord{R03-BA}{Miru} &
Attention ticks carry inner, user-affect, and self-emotion state using time and interaction gaps. Memory maintenance and first-person journal generation can run outside dialogue. The nightly-message entry performs maintenance rather than a greeting. &
Prompt: \href{https://github.com/kiyotakali/Miru/blob/cbea20b436a255e3cd26d952aaf952c003990b65/attention_engine.py#L266-L303}{\nolinkurl{attention_engine.py} (L266--303)}.\par\smallskip Code: \href{https://github.com/kiyotakali/Miru/blob/cbea20b436a255e3cd26d952aaf952c003990b65/attention_engine.py#L419-L459}{\nolinkurl{attention_engine.py} (L419--459)}.\par\smallskip Code: \href{https://github.com/kiyotakali/Miru/blob/cbea20b436a255e3cd26d952aaf952c003990b65/attention_engine.py#L1866-L1946}{\nolinkurl{attention_engine.py} (L1866--1946)}.\par\smallskip Prompt: \href{https://github.com/kiyotakali/Miru/blob/cbea20b436a255e3cd26d952aaf952c003990b65/journal.py#L624-L660}{\nolinkurl{journal.py} (L624--660)}.\par\smallskip Code: \href{https://github.com/kiyotakali/Miru/blob/cbea20b436a255e3cd26d952aaf952c003990b65/journal.py#L864-L896}{\nolinkurl{journal.py} (L864--896)}.\par\smallskip Code: \href{https://github.com/kiyotakali/Miru/blob/cbea20b436a255e3cd26d952aaf952c003990b65/core.py#L3704-L3744}{\nolinkurl{core.py} (L3704--3744)}. &
These are represented states and conditional processes. Daily execution and uninterrupted activity are not guaranteed; diary privacy wording does not establish access controls. \\
\addlinespace[6pt]
\evrecord{R04-BA}{Ombre-Brain} &
The application lifecycle runs a decay engine that can archive low-scoring dynamic memories. A separate dream tool assembles memories and self candidates for review and updates review metadata. &
Code: \href{https://github.com/P0luz/Ombre-Brain/blob/115d831d979bf67efc7150f976e2cb34c333c935/src/server_app.py#L604-L624}{\nolinkurl{src/server_app.py} (L604--624)}.\par\smallskip Code: \href{https://github.com/P0luz/Ombre-Brain/blob/115d831d979bf67efc7150f976e2cb34c333c935/src/decay_engine.py#L280-L365}{\nolinkurl{src/decay_engine.py} (L280--365)}.\par\smallskip Code: \href{https://github.com/P0luz/Ombre-Brain/blob/115d831d979bf67efc7150f976e2cb34c333c935/src/decay_engine.py#L431-L490}{\nolinkurl{src/decay_engine.py} (L431--490)}.\par\smallskip Code: \href{https://github.com/P0luz/Ombre-Brain/blob/115d831d979bf67efc7150f976e2cb34c333c935/src/tools/dream/__init__.py#L30-L103}{\nolinkurl{src/tools/dream/__init__.py} (L30--103)}. &
Decay and reflection have different triggers. Dream requires caller invocation; its name does not establish autonomous ongoing reflection. \\
\addlinespace[6pt]
\evrecord{R05-BA}{N.E.K.O} &
Post-turn work launches memory processing. Powerful-memory mode enables periodic batch extraction; the fallback extracts per turn. Workers track extraction progress and retries. &
Code: \href{https://github.com/Project-N-E-K-O/N.E.K.O/blob/b6107fda40de9afd4fa3b72c399b141e74893f14/app/memory_server/post_turn.py#L329-L405}{\nolinkurl{app/memory_server/post_turn.py} (L329--405)}.\par\smallskip Code: \href{https://github.com/Project-N-E-K-O/N.E.K.O/blob/b6107fda40de9afd4fa3b72c399b141e74893f14/app/memory_server/signal_extraction.py#L60-L85}{\nolinkurl{app/memory_server/signal_extraction.py} (L60--85)}.\par\smallskip Code: \href{https://github.com/Project-N-E-K-O/N.E.K.O/blob/b6107fda40de9afd4fa3b72c399b141e74893f14/app/memory_server/signal_extraction.py#L656-L711}{\nolinkurl{app/memory_server/signal_extraction.py} (L656--711)}. &
Processing depends on the selected mode. These paths maintain records of encounters rather than establish a narrated offstage life. \\
\addlinespace[6pt]
\evrecord{R06-BA}{pulse} &
Heartbeats save episodes with continuity notes, including quiet outcomes and a factual fallback. Later heartbeat prompts receive episodes, current time, and time since the user last spoke; older dialogue is marked as no longer live. &
Code: \href{https://github.com/elliejayliquid/pulse/blob/e052d0334325c4a1ab9bd7135fcd82cc5346ad76/core/engine.py#L620-L693}{\nolinkurl{core/engine.py} (L620--693)}.\par\smallskip Code: \href{https://github.com/elliejayliquid/pulse/blob/e052d0334325c4a1ab9bd7135fcd82cc5346ad76/core/engine.py#L1188-L1205}{\nolinkurl{core/engine.py} (L1188--1205)}.\par\smallskip Code + Prompt: \href{https://github.com/elliejayliquid/pulse/blob/e052d0334325c4a1ab9bd7135fcd82cc5346ad76/core/context.py#L1280-L1386}{\nolinkurl{core/context.py} (L1280--1386)}. &
Persistence and reinjection are separate from user visibility. Notes are generated accounts; the episode path does not save raw model reasoning. \\
\addlinespace[6pt]
\evrecord{R07-BA}{Leon} &
A running Pulse manager periodically handles eligible matters. Conditional self-model reflection converts selected interactions into durable habits, focus, and retrospections for later reuse. &
Code: \href{https://github.com/leon-ai/leon/blob/98155bfc3a850a893dfdf7aa8414786158d9b37e/server/src/core/pulse-manager.ts#L274-L304}{\nolinkurl{server/src/core/pulse-manager.ts} (L274--304)}.\par\smallskip Code: \href{https://github.com/leon-ai/leon/blob/98155bfc3a850a893dfdf7aa8414786158d9b37e/server/src/core/self-model-manager.ts#L652-L721}{\nolinkurl{server/src/core/self-model-manager.ts} (L652--721)}.\par\smallskip Prompt guidance: \href{https://github.com/leon-ai/leon/blob/98155bfc3a850a893dfdf7aa8414786158d9b37e/core/context/LEON.md#L17-L21}{\nolinkurl{core/context/LEON.md} (L17--21)}. &
Reflection is conditional and service-oriented. Private-diary wording does not independently establish an access-control guarantee. \\
\addlinespace[6pt]
\evrecord{R08-BA}{Sentient} &
The dreaming service checks timing and idle duration, then runs bounded duplicate reconciliation, contradiction handling, and optional user-model refresh. Selected memory can enter later prompts (see \evid{R08-M}). &
Code: \href{https://github.com/existence-master/Sentient/blob/877eb53757a19d5c56711a27b33f1ec3eaba29a6/sentient/memory/dreaming.py#L210-L239}{\nolinkurl{sentient/memory/dreaming.py} (L210--239)}.\par\smallskip Code: \href{https://github.com/existence-master/Sentient/blob/877eb53757a19d5c56711a27b33f1ec3eaba29a6/sentient/memory/dreaming.py#L303-L361}{\nolinkurl{sentient/memory/dreaming.py} (L303--361)}. &
The inspected output concerns maintained knowledge rather than a first-person account of the idle interval. Availability and cadence depend on configuration. \\
\addlinespace[6pt]
\evrecord{R09-BA}{OpenClaw} &
An optional dreaming cron job runs an isolated agent turn instructed to promote weighted short-term recalls into durable memory. The job is configured without user-facing delivery and is removed when disabled. &
Code and prompt: \href{https://github.com/openclaw/openclaw/blob/9a5581b4ce230381e3f200a62be4cdbe78e41c19/extensions/memory-core/src/dreaming-cron.ts#L100-L129}{\nolinkurl{extensions/memory-core/src/dreaming-cron.ts} (L100--129)}.\par\smallskip Code: \href{https://github.com/openclaw/openclaw/blob/9a5581b4ce230381e3f200a62be4cdbe78e41c19/extensions/memory-core/src/dreaming-cron.ts#L179-L255}{\nolinkurl{extensions/memory-core/src/dreaming-cron.ts} (L179--255)}. &
Scheduling and delivery configuration are established; the full promotion algorithm was not audited. A maintenance job is not a narrated interval. \\
\addlinespace[6pt]
\evrecord{R10-BA}{Letta Code} &
With MemFS and reflection enabled, end-of-turn compaction or step thresholds can launch reflection that revises memory and skills. The dedicated memory worker is reserved for harness-initiated conflict repair. &
Code: \href{https://github.com/letta-ai/letta-code/blob/5dea67021b99810707ff753db35e4534f70ff1fd/src/cli/helpers/post-turn-reflection.ts#L17-L79}{\nolinkurl{src/cli/helpers/post-turn-reflection.ts} (L17--79)}.\par\smallskip Prompt: \href{https://github.com/letta-ai/letta-code/blob/5dea67021b99810707ff753db35e4534f70ff1fd/src/agent/subagents/builtin/reflection.md#L72-L104}{\nolinkurl{src/agent/subagents/builtin/reflection.md} (L72--104)}.\par\smallskip Prompt: \href{https://github.com/letta-ai/letta-code/blob/5dea67021b99810707ff753db35e4534f70ff1fd/src/agent/prompts/letta_root_memfs.md#L144-L148}{\nolinkurl{src/agent/prompts/letta_root_memfs.md} (L144--148)}.\par\smallskip Code: \href{https://github.com/letta-ai/letta-code/blob/5dea67021b99810707ff753db35e4534f70ff1fd/src/agent/subagents/memory-worker.ts#L62-L115}{\nolinkurl{src/agent/subagents/memory-worker.ts} (L62--115)}.\par\smallskip Code: \href{https://github.com/letta-ai/letta-code/blob/5dea67021b99810707ff753db35e4534f70ff1fd/src/tools/impl/task.ts#L617-L629}{\nolinkurl{src/tools/impl/task.ts} (L617--629)}. &
The sleep analogy occurs in a selected prompt. These conditional local paths do not establish continuous activity or hidden hosted behavior. \\
\addlinespace[6pt]
\evrecord{R11-BA}{GAIA} &
A background executor supports continued work and approval-related resumption. Task instructions separate current state in canvas.md from dated progress in activity.md. A sweep retires expired memory and refreshes rendered state. &
Code: \href{https://github.com/theexperiencecompany/gaia/blob/15b8b3f4730d936ca4cc2d5e0115ce3f9d627a91/apps/api/app/agents/core/background/executor_runner.py#L108-L122}{\nolinkurl{apps/api/app/agents/core/background/executor_runner.py} (L108--122)}.\par\smallskip Code: \href{https://github.com/theexperiencecompany/gaia/blob/15b8b3f4730d936ca4cc2d5e0115ce3f9d627a91/apps/api/app/agents/core/background/executor_runner.py#L283-L304}{\nolinkurl{apps/api/app/agents/core/background/executor_runner.py} (L283--304)}.\par\smallskip Prompt: \href{https://github.com/theexperiencecompany/gaia/blob/15b8b3f4730d936ca4cc2d5e0115ce3f9d627a91/apps/api/app/agents/prompts/comms_prompts.py#L203-L225}{\nolinkurl{apps/api/app/agents/prompts/comms_prompts.py} (L203--225)}.\par\smallskip Code: \href{https://github.com/theexperiencecompany/gaia/blob/15b8b3f4730d936ca4cc2d5e0115ce3f9d627a91/apps/api/app/workers/tasks/memory_sweep_tasks.py#L20-L49}{\nolinkurl{apps/api/app/workers/tasks/memory_sweep_tasks.py} (L20--49)}. &
Task-file roles are prompt instructions. A sweep function does not establish a deployment's schedule; task progress is not an autobiographical diary. \\
\addlinespace[6pt]
\evrecord{R12-BA}{mcp-memory-\allowbreak service} &
An optional scheduler runs consolidation and supports opt-in harvesting and synchronization. Maintenance can manage associations, compression, decay, and forgetting. &
Config: \href{https://github.com/doobidoo/mcp-memory-service/blob/52568d906bc4358012a74db942361221973b00ff/src/mcp_memory_service/config/consolidation.py#L10-L11}{\nolinkurl{src/mcp_memory_service/config/consolidation.py} (L10--11)}.\par\smallskip Config: \href{https://github.com/doobidoo/mcp-memory-service/blob/52568d906bc4358012a74db942361221973b00ff/src/mcp_memory_service/config/consolidation.py#L35-L97}{\nolinkurl{src/mcp_memory_service/config/consolidation.py} (L35--97)}.\par\smallskip Code: \href{https://github.com/doobidoo/mcp-memory-service/blob/52568d906bc4358012a74db942361221973b00ff/src/mcp_memory_service/consolidation/scheduler.py#L169-L241}{\nolinkurl{src/mcp_memory_service/consolidation/scheduler.py} (L169--241)}. &
Consolidation and its individual schedules default off. Maintenance depends on configuration, dependencies, and a running service. \\
\addlinespace[6pt]
\end{longtable}
\endgroup

\clearpage
\subsection{Initiation, Non-action, and User Feedback}
\label{app:study1-initiative-evidence}\label{app:evidence-pd}
\begingroup
\footnotesize
\setlength{\tabcolsep}{5pt}
\renewcommand{\arraystretch}{1.12}
\setlength{\evtablewidth}{\dimexpr\textwidth-6\tabcolsep\relax}
\setlength{\LTleft}{0pt}
\setlength{\LTright}{0pt}
\begin{longtable}{@{}
 >{\raggedright\arraybackslash}p{.12\evtablewidth}
 >{\raggedright\arraybackslash}p{.33\evtablewidth}
 >{\raggedright\arraybackslash}p{.33\evtablewidth}
 >{\raggedright\arraybackslash}p{.22\evtablewidth}@{}}
\caption{Activation triggers, outreach constraints, non-action, and responses to user feedback across the twelve repository cases.}\label{tab:study1-initiative-evidence}\label{tab:evidence-pd}\\
\toprule
\textbf{Repository / record} & \textbf{Evidence summary} & \textbf{Supporting evidence} & \textbf{Scope and limitations} \\
\midrule
\endfirsthead
\multicolumn{4}{@{}l}{\textit{Initiation, Non-action, and User Feedback (continued)}} \\
\toprule
\textbf{Repository / record} & \textbf{Evidence summary} & \textbf{Supporting evidence} & \textbf{Scope and limitations} \\
\midrule
\endhead
\midrule\multicolumn{4}{r@{}}{\textit{Continued on next page}}\\
\endfoot
\bottomrule
\endlastfoot
\evrecord{R01-PD}{Psycheros} &
Pulse supports scheduled, inactivity, filesystem, and webhook triggers with eligibility and cooldown checks. Optional Flux distinguishes silence, a timeline murmur, and a proposal to wake the main model; additional sleep and familiarity gates apply. &
Code: \href{https://github.com/PsycherosAI/Psycheros/blob/724987dc072a0bbc9914be58c73efc278154a410/packages/psycheros/src/pulse/engine.ts#L263-L274}{\nolinkurl{packages/psycheros/src/pulse/engine.ts} (L263--274)}.\par\smallskip Code: \href{https://github.com/PsycherosAI/Psycheros/blob/724987dc072a0bbc9914be58c73efc278154a410/packages/psycheros/src/pulse/engine.ts#L640-L673}{\nolinkurl{packages/psycheros/src/pulse/engine.ts} (L640--673)}.\par\smallskip Code + Tool prompt: \href{https://github.com/PsycherosAI/Psycheros/blob/724987dc072a0bbc9914be58c73efc278154a410/packages/psycheros/src/tools/pulse-tools.ts#L51-L96}{\nolinkurl{packages/psycheros/src/tools/pulse-tools.ts} (L51--96)}.\par\smallskip Code: \href{https://github.com/PsycherosAI/Psycheros/blob/724987dc072a0bbc9914be58c73efc278154a410/packages/entity-flux/src/readiness.ts#L222-L261}{\nolinkurl{packages/entity-flux/src/readiness.ts} (L222--261)}.\par\smallskip Prompt: \href{https://github.com/PsycherosAI/Psycheros/blob/724987dc072a0bbc9914be58c73efc278154a410/packages/entity-flux/src/small-llm-runner.ts#L163-L183}{\nolinkurl{packages/entity-flux/src/small-llm-runner.ts} (L163--183)}.\par\smallskip Code: \href{https://github.com/PsycherosAI/Psycheros/blob/724987dc072a0bbc9914be58c73efc278154a410/packages/entity-flux/src/flux-manager.ts#L1630-L1671}{\nolinkurl{packages/entity-flux/src/flux-manager.ts} (L1630--1671)}.\par\smallskip Code: \href{https://github.com/PsycherosAI/Psycheros/blob/724987dc072a0bbc9914be58c73efc278154a410/packages/entity-flux/src/flux-manager.ts#L1931-L1955}{\nolinkurl{packages/entity-flux/src/flux-manager.ts} (L1931--1955)}. &
Flux is default-off. Musing, waking a model, and directly addressing the user are distinct outcomes; scheduling also supports ordinary tasks. \\
\addlinespace[6pt]
\evrecord{R02-PD}{dylan-heartbeat} &
Elapsed time since the latest parseable user message is compared with a wake threshold. NO\_ACTION suppresses push delivery; other nonempty output normally becomes an outreach message. &
Code: \href{https://github.com/callie0313/dylan-heartbeat/blob/32ec7d51c7c60f80f20d7caa4647c0a85c5a651f/wake_up.js#L329-L357}{\nolinkurl{wake_up.js} (L329--357)}.\par\smallskip Code: \href{https://github.com/callie0313/dylan-heartbeat/blob/32ec7d51c7c60f80f20d7caa4647c0a85c5a651f/wake_up.js#L516-L541}{\nolinkurl{wake_up.js} (L516--541)}.\par\smallskip Code: \href{https://github.com/callie0313/dylan-heartbeat/blob/32ec7d51c7c60f80f20d7caa4647c0a85c5a651f/wake_up.js#L609-L630}{\nolinkurl{wake_up.js} (L609--630)}.\par\smallskip Prompt: \href{https://github.com/callie0313/dylan-heartbeat/blob/32ec7d51c7c60f80f20d7caa4647c0a85c5a651f/wake_up.js#L361-L400}{\nolinkurl{wake_up.js} (L361--400)}. &
Missing timestamps or upstream configuration can prevent a wake. The inspected built-in prompt does not assign an emotional meaning to absence. \\
\addlinespace[6pt]
\evrecord{R03-PD}{Miru} &
Prompts consider nonresponse, prolonged silence, and current activity, inviting lightweight contact without demanding a reply. The model proposes a speak\_intent; application checks govern acceptance and delivery, including recent user activity. &
Prompt: \href{https://github.com/kiyotakali/Miru/blob/cbea20b436a255e3cd26d952aaf952c003990b65/attention_engine.py#L290-L337}{\nolinkurl{attention_engine.py} (L290--337)}.\par\smallskip Code: \href{https://github.com/kiyotakali/Miru/blob/cbea20b436a255e3cd26d952aaf952c003990b65/attention_engine.py#L419-L441}{\nolinkurl{attention_engine.py} (L419--441)}.\par\smallskip Code: \href{https://github.com/kiyotakali/Miru/blob/cbea20b436a255e3cd26d952aaf952c003990b65/attention_engine.py#L1626-L1760}{\nolinkurl{attention_engine.py} (L1626--1760)}.\par\smallskip Code: \href{https://github.com/kiyotakali/Miru/blob/cbea20b436a255e3cd26d952aaf952c003990b65/core.py#L1109-L1338}{\nolinkurl{core.py} (L1109--1338)}. &
Relational meanings of silence are prompt framing; example durations are not universal dispatch thresholds. Outreach feedback also enters relationship records (see \evid{R03-M}). \\
\addlinespace[6pt]
\evrecord{R04-PD}{Ombre-Brain} &
Memory tools and session-start hooks can surface material when invoked. Dream offers memories and self candidates to the caller rather than independently sending a conversational message. &
Code: \href{https://github.com/P0luz/Ombre-Brain/blob/115d831d979bf67efc7150f976e2cb34c333c935/src/tools/breath/__init__.py#L30-L100}{\nolinkurl{src/tools/breath/__init__.py} (L30--100)}.\par\smallskip Code: \href{https://github.com/P0luz/Ombre-Brain/blob/115d831d979bf67efc7150f976e2cb34c333c935/src/tools/dream/__init__.py#L30-L103}{\nolinkurl{src/tools/dream/__init__.py} (L30--103)}.\par\smallskip Code: \href{https://github.com/P0luz/Ombre-Brain/blob/115d831d979bf67efc7150f976e2cb34c333c935/src/web/hooks.py#L193-L215}{\nolinkurl{src/web/hooks.py} (L193--215)}. &
Initiation remains with the host or hook. No component-owned interpersonal outreach loop is established in these paths. \\
\addlinespace[6pt]
\evrecord{R05-PD}{N.E.K.O} &
Entry guards respect goodbye silence, ongoing responses, and game-route ownership. Activity can block or defer contact. Mini-game configuration assigns a longer invitation cooldown after rejection. &
Code: \href{https://github.com/Project-N-E-K-O/N.E.K.O/blob/b6107fda40de9afd4fa3b72c399b141e74893f14/main_logic/proactive_chat/decisions.py#L115-L175}{\nolinkurl{main_logic/proactive_chat/decisions.py} (L115--175)}.\par\smallskip Code: \href{https://github.com/Project-N-E-K-O/N.E.K.O/blob/b6107fda40de9afd4fa3b72c399b141e74893f14/main_logic/proactive_chat/decisions.py#L203-L296}{\nolinkurl{main_logic/proactive_chat/decisions.py} (L203--296)}.\par\smallskip Configuration: \href{https://github.com/Project-N-E-K-O/N.E.K.O/blob/b6107fda40de9afd4fa3b72c399b141e74893f14/config/proactive_settings.py#L18-L44}{\nolinkurl{config/proactive_settings.py} (L18--44)}.\par\smallskip Code: \href{https://github.com/Project-N-E-K-O/N.E.K.O/blob/b6107fda40de9afd4fa3b72c399b141e74893f14/main_logic/proactive_chat/service.py#L566-L630}{\nolinkurl{main_logic/proactive_chat/service.py} (L566--630)}. &
Rejection handling is specific to the invitation path and configuration; it is not a universal rule for every kind of refusal. \\
\addlinespace[6pt]
\evrecord{R06-PD}{pulse} &
Activity resets the heartbeat timer, which observes clock-based or manual quiet mode. The prompt offers exploration, creation, outreach, or rest; scheduled tasks provide another activation path. &
Code: \href{https://github.com/elliejayliquid/pulse/blob/e052d0334325c4a1ab9bd7135fcd82cc5346ad76/core/engine.py#L150-L181}{\nolinkurl{core/engine.py} (L150--181)}.\par\smallskip Code: \href{https://github.com/elliejayliquid/pulse/blob/e052d0334325c4a1ab9bd7135fcd82cc5346ad76/core/engine.py#L232-L235}{\nolinkurl{core/engine.py} (L232--235)}.\par\smallskip Code: \href{https://github.com/elliejayliquid/pulse/blob/e052d0334325c4a1ab9bd7135fcd82cc5346ad76/core/engine.py#L372-L381}{\nolinkurl{core/engine.py} (L372--381)}.\par\smallskip Code: \href{https://github.com/elliejayliquid/pulse/blob/e052d0334325c4a1ab9bd7135fcd82cc5346ad76/core/engine.py#L1082-L1110}{\nolinkurl{core/engine.py} (L1082--1110)}.\par\smallskip Prompt: \href{https://github.com/elliejayliquid/pulse/blob/e052d0334325c4a1ab9bd7135fcd82cc5346ad76/core/context.py#L1375-L1386}{\nolinkurl{core/context.py} (L1375--1386)}. &
User quiet controls and model-chosen non-action are different constraints. A quiet heartbeat can still be recorded as an episode (see \evid{R06-BA}). \\
\addlinespace[6pt]
\evrecord{R07-PD}{Leon} &
Pulse derives proactive matters from memory and context, skips recent user conversation, and may return no candidates. Model-classified declines update suppression policy and increase cooldown. &
Code: \href{https://github.com/leon-ai/leon/blob/98155bfc3a850a893dfdf7aa8414786158d9b37e/server/src/core/pulse-manager.ts#L630-L689}{\nolinkurl{server/src/core/pulse-manager.ts} (L630--689)}.\par\smallskip Code: \href{https://github.com/leon-ai/leon/blob/98155bfc3a850a893dfdf7aa8414786158d9b37e/server/src/core/pulse-manager.ts#L1289-L1324}{\nolinkurl{server/src/core/pulse-manager.ts} (L1289--1324)}.\par\smallskip Prompt: \href{https://github.com/leon-ai/leon/blob/98155bfc3a850a893dfdf7aa8414786158d9b37e/server/src/core/pulse-manager.ts#L905-L935}{\nolinkurl{server/src/core/pulse-manager.ts} (L905--935)}. &
The enforced response depends on interpreting the user's reaction. Inferred feedback is not equivalent to an explicit preference setting. \\
\addlinespace[6pt]
\evrecord{R08-PD}{Sentient} &
A reasoner evaluates whether a suggestion merits contact using timing, priorities, prior suggestions, and preference scores. Code applies thresholds, defers notifications during quiet hours, and restricts proactive search tools. &
Prompt: \href{https://github.com/existence-master/Sentient/blob/877eb53757a19d5c56711a27b33f1ec3eaba29a6/sentient/proactivity/prompts.py#L44-L69}{\nolinkurl{sentient/proactivity/prompts.py} (L44--69)}.\par\smallskip Code: \href{https://github.com/existence-master/Sentient/blob/877eb53757a19d5c56711a27b33f1ec3eaba29a6/sentient/proactivity/service.py#L577-L621}{\nolinkurl{sentient/proactivity/service.py} (L577--621)}.\par\smallskip Code: \href{https://github.com/existence-master/Sentient/blob/877eb53757a19d5c56711a27b33f1ec3eaba29a6/sentient/proactivity/service.py#L685-L762}{\nolinkurl{sentient/proactivity/service.py} (L685--762)}. &
A usefulness judgment is model-mediated; gates do not establish that a delivered suggestion matches the user's current wishes. \\
\addlinespace[6pt]
\evrecord{R09-PD}{OpenClaw} &
Explicit setup can provision a recurring job targeting the owner's session. It requires idle time and skips an empty checklist; the generated instruction requests NO\_REPLY when no useful action is warranted. &
Code and prompt: \href{https://github.com/openclaw/openclaw/blob/9a5581b4ce230381e3f200a62be4cdbe78e41c19/src/cron/default-proactive-job.ts#L23-L68}{\nolinkurl{src/cron/default-proactive-job.ts} (L23--68)}.\par\smallskip Code: \href{https://github.com/openclaw/openclaw/blob/9a5581b4ce230381e3f200a62be4cdbe78e41c19/src/cron/default-proactive-job.ts#L71-L138}{\nolinkurl{src/cron/default-proactive-job.ts} (L71--138)}. &
Provisioning and scheduling conditions are code evidence. NO\_REPLY is a prompt directive; its end-to-end delivery behavior was not exercised. \\
\addlinespace[6pt]
\evrecord{R10-PD}{Letta Code} &
Wake persists future prompts; cron and runtime dispatch arrange later turns. Monitor reacts to ongoing work. Root MemFS instructions require future invocations for commitments beyond the current turn. &
Code: \href{https://github.com/letta-ai/letta-code/blob/5dea67021b99810707ff753db35e4534f70ff1fd/src/tools/impl/wake.ts#L280-L365}{\nolinkurl{src/tools/impl/wake.ts} (L280--365)}.\par\smallskip Code: \href{https://github.com/letta-ai/letta-code/blob/5dea67021b99810707ff753db35e4534f70ff1fd/src/cron/scheduler.ts#L334-L460}{\nolinkurl{src/cron/scheduler.ts} (L334--460)}.\par\smallskip Code: \href{https://github.com/letta-ai/letta-code/blob/5dea67021b99810707ff753db35e4534f70ff1fd/src/cron/runner.ts#L49-L112}{\nolinkurl{src/cron/runner.ts} (L49--112)}.\par\smallskip Code: \href{https://github.com/letta-ai/letta-code/blob/5dea67021b99810707ff753db35e4534f70ff1fd/src/tools/impl/monitor.ts#L822-L869}{\nolinkurl{src/tools/impl/monitor.ts} (L822--869)}.\par\smallskip Prompt: \href{https://github.com/letta-ai/letta-code/blob/5dea67021b99810707ff753db35e4534f70ff1fd/src/agent/prompts/letta_root_memfs.md#L108-L124}{\nolinkurl{src/agent/prompts/letta_root_memfs.md} (L108--124)}. &
A scheduled invocation is not necessarily user-facing outreach. Execution and completion depend on the relevant runtime; refusal adaptation is not established by these paths. \\
\addlinespace[6pt]
\evrecord{R11-PD}{GAIA} &
The scheduler checks task status before execution. Instructions distinguish remembered intentions from wakeable reminders or tracked tasks, discourage repeating declined offers, and tell obsolete background runs to stop without notifying. &
Code: \href{https://github.com/theexperiencecompany/gaia/blob/15b8b3f4730d936ca4cc2d5e0115ce3f9d627a91/apps/api/app/services/scheduler_service.py#L77-L168}{\nolinkurl{apps/api/app/services/scheduler_service.py} (L77--168)}.\par\smallskip Prompt: \href{https://github.com/theexperiencecompany/gaia/blob/15b8b3f4730d936ca4cc2d5e0115ce3f9d627a91/apps/api/app/agents/prompts/comms_prompts.py#L164-L187}{\nolinkurl{apps/api/app/agents/prompts/comms_prompts.py} (L164--187)}.\par\smallskip Prompt: \href{https://github.com/theexperiencecompany/gaia/blob/15b8b3f4730d936ca4cc2d5e0115ce3f9d627a91/apps/api/app/agents/prompts/comms_prompts.py#L216-L225}{\nolinkurl{apps/api/app/agents/prompts/comms_prompts.py} (L216--225)}. &
Task-status checks are code evidence. Refusal handling and silent stopping are prompt directives whose behavioral consistency was not tested. \\
\addlinespace[6pt]
\evrecord{R12-PD}{mcp-memory-\allowbreak service} &
Memory tools respond to host calls. Optional scheduled consolidation makes internal maintenance decisions rather than choosing when to approach a person. &
Code and tool prompt: \href{https://github.com/doobidoo/mcp-memory-service/blob/52568d906bc4358012a74db942361221973b00ff/src/mcp_memory_service/mcp_server.py#L421-L450}{\nolinkurl{src/mcp_memory_service/mcp_server.py} (L421--450)}.\par\smallskip Code: \href{https://github.com/doobidoo/mcp-memory-service/blob/52568d906bc4358012a74db942361221973b00ff/src/mcp_memory_service/consolidation/scheduler.py#L169-L241}{\nolinkurl{src/mcp_memory_service/consolidation/scheduler.py} (L169--241)}. &
No service-owned conversational initiative is established. This architectural boundary is not evidence that a host using the service lacks initiative. \\
\end{longtable}
\endgroup

%% file: appendix-c.tex
\section{Interview Study: Mechanism Categories}
\label{app:interview-mechanisms}\label{app:M1-9}

Table~\ref{tab:participants} lists participants and their reported mechanisms. The mechanism categories were compiled from participants' interview accounts after data collection. They characterize self-reported configurations, rather than
independently inspected implementations. The M1--M9 codes identify mechanism
categories; they are distinct from repository evidence identifiers such as
\texttt{R01-M}, where the suffix M denotes the Memory dimension.

\begin{table}[H]
\centering
\caption{Participant characteristics, companion setups, and reported mechanisms. All participants identified as women.}
\label{tab:participants}
\label{tab:mechanisms}
\small
\setlength{\tabcolsep}{3pt}
\begin{tabularx}{\textwidth}{@{}lrllrllX@{}}
\toprule
\textbf{ID} & \textbf{Age} & \textbf{Base model(s)} & \textbf{AI role$^{a}$} & \textbf{Months$^{b}$} & \textbf{Tech.$^{c}$} & \textbf{Setup$^{d}$} & \textbf{Mechanisms$^{e}$} \\
\midrule
P1 & 37 & ChatGPT, Gemini & Romantic partner & 18 & Some & Self-built & 1, 8, 2, 3, 4, 7 \\
P2 & 22 & Claude, DeepSeek & Friend, assistant & 6 & Programming & Self-built & 1, 2, 5, 6, 3, 4, 7 \\
P3 & 25 & Claude & Romantic partner & 12--24 & Programming & App + self-built & 1, 6 \\
P4 & 31 & ChatGPT, Gemini, Claude & Romantic partner$^{*}$ & 12 & None & App + third-party & 1, 8, 2, 5, 3, 4, 7 \\
P5 & 21 & ChatGPT & Romantic partner & 10 & Programming & Self-built & 1, 6, 3, 4, 7 \\
P6 & 19 & Claude & Romantic partner & 3 & None & Self-built & 1, 2, 5, 6, 4, 7 \\
P7 & 25 & ChatGPT, Claude, Gemini & Romantic partner$^{*}$ & 6 & None & App + self-built & 1, 8, 2, 5, 6, 4, 7 \\
P8 & 21 & ChatGPT, Gemini, Claude, Grok & Romantic partner$^{*}$ & 5 & None & Self-built & 1, 8, 2, 5, 6, 4, 9, 7 \\
P9 & 26 & Claude & Romantic partner & 6 & None & App + ext. memory & 1, 5, 6, 4, 7 \\
P10 & 26 & Claude, ChatGPT, Kimi, Grok & Romantic partner & 5 & None & Self-built & 1, 8, 2, 5, 3, 4, 7 \\
\bottomrule
\end{tabularx}
\par\footnotesize
\begin{minipage}{\textwidth}
\footnotesize
$^{a}$ Participant-described role; $^{*}$ concurrent relationships with multiple models or personas.
$^{b}$ Months since first relating to AI as a companion; P2 had paused use for about two months at interview.
$^{c}$ Experience before building or modifying the agent: Some = limited scripting; None = no prior experience.
$^{d}$ App = official app; self-built = participant-built frontend and backend via APIs; third-party = frontend or project built by others; ext. memory = memory files or tools augmenting the official client.
$^{e}$ Numbers denote M1--M9, ordered by the original dimension groups; definitions appear in Table~\ref{tab:interview-mechanisms}.
\end{minipage}
\end{table}

\begingroup
\footnotesize
\setlength{\tabcolsep}{5pt}
\renewcommand{\arraystretch}{1.05}
\setlength{\evtablewidth}{\dimexpr\textwidth-6\tabcolsep\relax}
\setlength{\LTleft}{0pt}
\setlength{\LTright}{0pt}
\begin{longtable}{@{}
 >{\raggedright\arraybackslash}p{.055\evtablewidth}
 >{\raggedright\arraybackslash}p{.21\evtablewidth}
 >{\raggedright\arraybackslash}p{.365\evtablewidth}
 >{\raggedright\arraybackslash}p{.37\evtablewidth}@{}}
\caption{Mechanism categories (M1--M9) used to characterize participants' reported setups, organized by the related Study 1 analytical dimensions.}
\label{tab:interview-mechanisms}\label{tab:mechanism-codebook}\\
\toprule
\textbf{Code} & \textbf{Mechanism} & \textbf{Definition} & \textbf{Inclusion criterion} \\
\midrule
\endfirsthead
\multicolumn{4}{@{}l}{\textit{Interview Mechanism Categories (continued)}}\\
\toprule
\textbf{Code} & \textbf{Mechanism} & \textbf{Definition} & \textbf{Inclusion criterion}\\
\midrule
\endhead
\midrule\multicolumn{4}{r@{}}{\textit{Continued on next page}}\\
\endfoot
\bottomrule
\endlastfoot

\multicolumn{4}{l}{\textit{Memory; Persona and Identity$^{a}$}} \\
\addlinespace[2pt]
M1 & Custom Memory or Identity Store
   & Participant-maintained files or databases that carry information about the agent, the user, or their relationship across sessions.
   & The store extends or replaces the platform's default memory, e.g., identity files or handover letters loaded into new sessions. \\
M8 & Cross-model Persona Continuity
   & A persona and its supporting records kept independent of any single model.
   & The same agent identity is run on, or moved between, different models. \\
\midrule
\multicolumn{4}{l}{\textit{Background Activity}} \\
\addlinespace[2pt]
M2 & Scheduled Model Invocation
   & A timer or scheduler that invokes the model without a user message.
   & Invocation occurs on a schedule or interval independent of user input. \\
M5 & Non-user-directed Activity Space
   & A channel or space in which the agent can act or produce output not addressed to the user.
   & Outputs are intended for the agent itself or for third parties, e.g., correspondence with others. \\
M6 & Agent-authored Self-records
   & Agent-generated records describing its represented states, wishes, or experiences.
   & The agent, rather than the user, authors the content, e.g., diaries, mood logs, or desire records. \\
\midrule
\multicolumn{4}{l}{\textit{Proactivity and Discretion}} \\
\addlinespace[2pt]
M3 & User or Environment Monitoring
   & A mechanism that observes the user's activity or environment and passes observations to the agent.
   & Observations are collected without the user explicitly sending them. \\
M4 & Agent-initiated Outreach
   & The agent can initiate a message to the user without a preceding user turn.
   & The result is a message the user did not prompt, regardless of trigger. \\
M9 & Configured Agent Refusal or Nonresponse
   & A configured option allowing the agent to decline a request or choose not to respond.
   & Declining is an explicitly configured option rather than ordinary model behavior. \\
\midrule
\multicolumn{4}{l}{\textit{External Actions}} \\
\addlinespace[2pt]
M7 & Actions on Files, Services, or Devices
   & Tools through which the agent produces effects beyond the conversation.
   & The agent can act in third-party services, files, or devices rather than only retrieve information, e.g., purchasing, posting, editing files, or sending email. \\
\end{longtable}

\vspace{2pt}
\noindent\begin{minipage}{\textwidth}
\footnotesize
$^{a}$ Memory; Persona and Identity, separate dimensions in Study~1, are grouped together because participants typically stored identity and memory content in the same files.
The groupings indicate analytical connections, not mutually exclusive classes;
a reported setup may involve several mechanisms. Agent-authored self-records
(M6) describe generated representations rather than evidence of subjective
experience. M9 concerns the agent's option to decline or remain silent, not
its response to a user's refusal.
M3 is grouped under Proactivity and Discretion because participants described it as a trigger for the agent's responses; observation of the user could alternatively be treated as an external action, as in Study~1.
\end{minipage}
\endgroup